\documentclass[10pt]{iopart}
\usepackage[english]{babel}
\usepackage[T1]{fontenc}
\usepackage{lmodern}
\usepackage[export]{adjustbox}
\usepackage{graphicx, wrapfig}
\usepackage{caption}
\usepackage{subcaption}
\usepackage{textcomp}
\usepackage{gensymb}
\usepackage{stfloats}
\usepackage{cuted}

\usepackage{booktabs, multirow}
\usepackage{makecell}
\usepackage{tabularx}
\usepackage[table]{xcolor}
\newcolumntype{L}[1]{>{\raggedright\let\newline\\\arraybackslash\hspace{0pt}}m{#1}}

\definecolor{airforceblue}{rgb}{0.36, 0.54, 0.66}
\definecolor{lightgray}{rgb}{0.89, 0.90, 0.873}
\definecolor{celadon}{rgb}{0.67, 0.88, 0.69} 
\definecolor{darkseagreen}{rgb}{0.56, 0.74, 0.56} 
\definecolor{caribbeangreen}{rgb}{0.0, 0.8, 0.6} 
\definecolor{emerald}{rgb}{0.31, 0.78, 0.47} 
\definecolor{goldenyellow}{rgb}{1.0, 0.87, 0.0} 
\definecolor{goldenpoppy}{rgb}{0.99, 0.76, 0.0}
\definecolor{deepsaffron}{rgb}{1.0, 0.6, 0.2} 
\definecolor{flame}{rgb}{0.89, 0.35, 0.13} 
\definecolor{jasper}{rgb}{0.84, 0.23, 0.24} 
\definecolor{harvardcrimson}{rgb}{0.79, 0.0, 0.09} 

\definecolor{barblue}{RGB}{153,204,254}
\definecolor{groupblue}{RGB}{51,102,254}
\definecolor{linkred}{RGB}{165,0,33}
\usepackage{pifont}
\usepackage{multicol}
\usepackage{pgfgantt}
\usepackage{array}
\usepackage{setspace}

\usepackage{xr}
\usepackage[hidelinks]{hyperref}
\hypersetup{colorlinks=false}
\usepackage{enumitem} 

\newcounter{todoCount}[section]
\def\todoCheck{
    \ifnum \arabic{todoCount}>0

    {\color{red}
    \textrmit{Unresolved TODO commands in above section: \arabic{todoCount}}
    }

    \fi
}

\usepackage[toc,page]{appendix}

\usepackage{csquotes}
\usepackage[sorting=none, eprint=false, url=false]{biblatex}
\AtBeginBibliography{\footnotesize}
\usepackage[colorinlistoftodos,prependcaption,textsize=tiny, textwidth=20mm]{todonotes}

        \newcommand{\thetavec}{\mathbf{\theta}}

\newcommand{\divmidas}{D-MIBAS}

\newcommand{\etemp}{$T_{\mathrm{e}}$}
\newcommand{\edens}{$n_{\mathrm{e}}$}
\newcommand{\ndens}{$n_{\mathrm{H}}$}
\newcommand{\qmol}{$Q_\mathrm{mol}$}
\newcommand{\hendens}{$n_{{\mathrm{He}}^{{0}}}$}
\newcommand{\hepdens}{$n_{{\mathrm{He}}^{{1+}}}$}

\newcommand{\sgn}{\mathrm{sgn}}

\newcommand{\pec}{\mathrm{PEC}}
\newcommand{\dtwoeff}{D{2,\mathrm{eff}}}

\newcommand{\logposterior}{$\mathcal{L}_{\mathrm{Posterior}}$}
\newcommand{\loglikelihood}{$\mathcal{L}_{\mathrm{Likelihood}}$}
\newcommand{\logprior}{$\mathcal{L}_{\mathrm{Prior}}$}

\newcommand{\jsat}{$j_{\mathrm{sat}}$}

\newcommand{\posterior}[1]{\romannumeral#1\relax}
\newcommand{\state}[1]{%
  \ifnum #1=1
    detached case%
  \fi
  \ifnum #1=2
    attached case%
  \fi
}
\newcommand{\standardform}[2]{\ensuremath{{#1}\times 10^{#2}}}
\begin{document}
\title[Advancements in \divmidas{}]{Two-dimensional inference of divertor plasma characteristics: advancements to a multi-instrument Bayesian analysis system}

\author{D Greenhouse$^1$$^2$, C Bowman$^2$, B Lipschultz$^1$, K Verhaegh$^2$$^3$, J Harrison$^2$, A. Fil$^4$}

\address{$^1$York Plasma Institute, School of Physics, Engineering and Technology, University of York, Heslington, York YO10~5DD, UK}

\address{$^2$ United Kingdom Atomic Energy Authority, Culham Campus, Abingdon, Oxfordshire, OX14~3DB, UK}

\address{$^3$ Department of Applied Physics and Science Education, Eindhoven University of Technology, Eindhoven 5600 MB, Netherlands}

\address{$^4$ CEA, IRFM, F-13108 Saint-Paul-lez-Durance, France}

\ead{daniel.greenhouse@ukaea.uk}

\begin{abstract}


An integrated data analysis system based on Bayesian inference has been developed for application to data from multiple diagnostics over the two-dimensional cross-section of tokamak divertors. The divertor multi-instrument Bayesian analysis system (\divmidas{}) has been tested on a synthetic dataset (including realistic experimental uncertainties) generated from SOLPS-ITER predictions of the MAST-U divertor. The resulting inference was within 6\%, 5\%, and 30\% median absolute percentage error of the SOLPS-ITER-predicted electron temperature, electron density, and neutral atomic hydrogen density, respectively, across a two-dimensional poloidal cross-section of the MAST-U Super-X outer divertor. 

To accommodate molecular contributions to Balmer emission, an advanced emission model has been developed. This is shown to be crucial for inference accuracy. Our \divmidas{} system utilises a mesh aligned to poloidal magnetic flux-surfaces, throughout the divertor, with plasma parameters assigned to each mesh vertex and collectively considered in the inference. This allowed comprehensive forward models of multiple diagnostics and the inclusion of expected physics. This is shown to be important for inference precision when including molecular contributions to Balmer emission. These developments pave the way for accurate, two-dimensional electron temperature, electron density, and neutral atomic hydrogen density inferences for MAST-U divertor experimental data for the first time.

\end{abstract}

\noindent{\it Keywords\/}: Bayesian inference, integrated data analysis, divertor physics, Balmer line emission, plasma-molecule interaction

\submitto{\PPCF}
\maketitle 
\ioptwocol

\section{Introduction}

Mitigating the intense heat flux, from the core plasma to the plasma-facing components, to keep it within tolerable engineering limits is a crucial challenge for the realisation of fusion energy \cite{Wenninger2014,Pitts2019}. This heat flux is diverted, in most current and planned magnetic confinement fusion reactor designs, towards a designated power exhaust region, the divertor. Within the divertor, complex two-dimensional (and occasionally three-dimensional) phenomena occur including: plasma-neutral interactions \cite{Verhaegh2019,Lipschultz_1997}; impurity radiation \cite{Henderson2018,Reimold2015,Bernert2023}; and plasma-molecular chemistry \cite{Verhaegh_2023,Verhaegh_2022,Verhaegh2021,Verhaegh2021b}. These interactions can lead to detachment of the divertor plasma from the divertor surface (target) by simultaneously inducing losses of power, momentum, and particles (i.e., ion losses) resulting in multiple-orders-of-magnitude heat flux reductions to the divertor targets \cite{Lipschultz_1997, Lipschultz2007a}.   

Our understanding of divertor physics will inform the design of the power exhaust for future, reactor-class devices \cite{Wenninger2014,Osawa2023,Hudoba2023,Kuang2020}. Novel diagnostic and analysis techniques that measure or infer plasma characteristics over the entire 2D cross-section of the divertor plasma would, ideally, be developed to enhance current understanding of divertor physics and to provide detailed validation of plasma-edge simulations \cite{WIESEN2015480} that are used to extrapolate current knowledge to planned devices. Achieving this requires either direct measurement or inference of underlying plasma parameters such as electron temperature (\etemp{}), electron density (\edens{}), and, neutral atomic hydrogen density (\ndens{}). 

Current procedures for extracting 2D profiles of \etemp{} and \ndens{} with sufficient spatial resolution ($\sim$1 cm \cite{Eich_2013}) are limited. Point-like measurements of \etemp{} and \edens{} provided by Langmuir Probes and Thomson scattering diagnostics have been propagated to 2D profiles via strike point sweeping (similar to \cite{Allen1997}). However, the constant plasma conditions and accurate equilibrium reconstruction required for this analysis limits its use and introduces significant uncertainty. Alternative approaches at JET \cite{Karhunen2020} and TCV \cite{Perek2019submitted,Perek_2022} have made use of multi-wavelength imaging camera systems of multiple hydrogen Balmer and helium lines. They utilise atomic data from the ADAS database \cite{adas} to link the measured line-brightness to \etemp{}, \edens{}, and \ndens{}. However, this is an `ill-posed' problem where multiple combinations of \etemp{}, \edens{}, and \ndens{} can produce the same measured brightness. This, along with the uncertainties associated with the required tomographic inversion of camera images, limits the precision of inferred \etemp{} and \ndens{} profiles in such approaches.

Integrated data analysis (IDA) offers a route to the inference of 2D profiles of underlying plasma characteristics with low uncertainties and adequate spatial resolution by collectively considering a group of diagnostics \cite{Bowman_2020}. Bayesian inference techniques offer a mechanism for integrated data analysis through the utilisation of `forward models' of expected diagnostic responses based on underlying plasma parameters. With these forward models, the plasma parameters that most plausibly gave rise to the data recorded by all available diagnostics can be inferred in a rigorous statistical framework. Diagnostics that directly or indirectly measure the underlying plasma characteristics can be included; the quantities of an indirect measurement (such as filtered imaging diagnostics and spectroscopy \cite{Perek_2022,Verhaegh_2022}) are connected to the underlying plasma parameters through the forward model. Through the use of synthetic data, \cite{Bowman_2020} demonstrated that this approach can result in an accurate inference of the two-dimensional poloidal cross sections of \etemp{}, \edens{}, and \ndens{}. Multi-wavelength imaging of emission from multiple hydrogen Balmer lines \cite{Perek_2022,Wijkamp2023}, enabled by novel diagnostics such as MANTIS on the TCV tokamak \cite{Perek2019submitted,Perek_2022} and the MWI on the MAST-U tokamak \cite{Feng2021,Wijkamp2023}, are particularly useful to such IDA techniques due to their coverage over the entire divertor cross-section. 

Recent work \cite{Verhaegh_2023,Verhaegh_2022,Verhaegh2021b,Perek_2022,Karhunen2023} has shown that plasma-molecule interactions (PMI), ultimately resulting in excited atoms that emit atomic line emission, can have a significant impact on the total emissivity of hydrogen Balmer lines \footnote{especially that of the Balmer-$\alpha$ (3$\to$2) line}. Accommodating PMI within a Balmer line emission forward model requires parameters in addition to the \etemp{}, \edens{}, and \ndens{} used to model Balmer emission in \cite{Bowman_2020} (which neglected PMI). These additional parameters can give rise to multiple conflicting inference \textit{solutions}: the same data corresponding to Balmer line emission can be explained by an unspecified proportion of PMI, further complicating the ill-posed problem. These conflicting solutions result in a degradation of the inference, potentially preventing its reliable use on experimental data.

In the MAST-U Super-X divertor, plasma-molecular chemistry has been shown to play a dominant role in hydrogen emission and particle exhaust \cite{Verhaegh_2023,Verhaegh_2022,Verhaegh2023c}. The impact of PMI on the hydrogen line emission appears to be particularly significant in strongly baffled divertors, as well as alternative divertor configurations (ADC) \cite{Verhaegh_2023,Verhaegh_2022,Verhaegh2023c}. `Neutral baffling' encloses most of the divertor region with surfaces allowing higher neutral and molecular densities \cite{Reimerdes2021}. For the same conditions as more conventional divertors, ADCs may permit more deeply detached conditions where there is greater loss of momentum, energy, and ions. Generating additional volume between the ionising (attached) region of the divertor plasma and the target permits even higher quantities of neutral atoms and molecules (and the accompanying particle, energy, and momentum sinks) \cite{Verhaegh_2023}. Ongoing experiments at MAST-U and TCV are focused on investigating such novel ADCs \cite{Moulton2023,Harrison2023,Verhaegh2023d,Theiler2017}. MAST-U has been designed to operate with a baffled `Super-X' configuration which is characterised by enhanced (compared to conventional configurations or other ADCs) total-flux expansion \cite{Lipschultz_2016}; the cross-sectional area of the divertor flux tubes increases from the x-point region to the outer divertor target. This is achieved by placing the divertor target at a much larger major radius than the x-point \cite{Havlickova2015}, increasing the ratio between the magnetic field at the x-point and the target \cite{Lipschultz_2016}. Consistent with recent experimental results from MAST-U \cite{Verhaegh_2023,Verhaegh2023c,Moulton2023,Harrison2023}, an increase in total flux expansion is expected to: reduce target heat loads; reduce upstream electron densities and divertor impurity fractions required for detachment; and reduce the sensitivity of the detached region length to changes in fuelling/impurity seeding \cite{Lipschultz_2016,Havlickova2015}. 

This work presents an improved framework that allows for accurate inferences (solutions) of \etemp{}, \edens{}, and \ndens{} across the entire divertor cross-section in both attached and detached Super-X regimes. Building on previous work \cite{Bowman_2020}, data from multiple diagnostics are combined and interpreted together using Bayesian inference. This work features filtered imaging diagnostic measurements of both helium singlet line emission and hydrogen Balmer line emission, whilst accounting for the contribution of PMI in the hydrogen Balmer line emission. By conducting the inference on a mesh (see Section~\ref{sec:cell vs mesh}) aligned to poloidal magnetic flux-surfaces, physics constraints (e.g. the expected monotonic reduction in static electron pressure along a flux tube to the target \cite{Stangeby1993}) can be included in the inference as prior knowledge. These `priors' are found to greatly improve the robustness of the analysis and reduce its inferred uncertainty, easing the ill-posed nature of the inference. Synthetic tests, performed by applying synthetic diagnostics with realistic experimental uncertainties to SOLPS-ITER simulations of the MAST-U Super-X divertor, have been performed to investigate the performance of this new Bayesian multi-diagnostic inference technique.

\section{Integrated Data Analysis Approaches}

\subsection{Bayesian Inference}
Bayesian inference is widely used in integrated data analysis, providing a robust statistical framework which inherently provides uncertainty quantification. 
Bayesian inference hypothesises that the data from each diagnostic, $\mathcal{D}$, are fundamentally certain. For each diagnostic, forward models are used to predict the data from model parameters, $\theta$, which (from a Bayesian perspective) are fundamentally uncertain. The \textit{posterior} probability of a particular combination of model parameters given these data is given by
\begin{equation}
    \underbrace{P(\theta|\mathcal{D})}_{\mathrm{posterior}} \propto  \underbrace{P(\mathcal{D}|\theta)}_{\mathrm{likelihood}}\times \underbrace{P(\theta)}_{\mathrm{prior}}.
    \label{eqn:bayes}
\end{equation}
The many different parameter combinations result in a posterior distribution. The mode of this distribution is the \emph{maximum a posteriori} (MAP) estimate which, in this work, is taken to be the inferred $\theta$ (solution). Sampling from the distribution provides the uncertainty on the inferred $\theta$. In this work, the uncertainty is taken to be the 95\% highest density interval (HDI)\footnote{95\% HDI is the most narrow interval that contains 95\% of the samples}. 

Accurate inference depends on the MAP estimate being correctly found and the forward models being accurate and comprehensive. The forward models specify which parameters, $\theta$, are required in the inference. To guarantee that the MAP estimate and the uncertainty are correctly found, all plausible parameter combinations (with sufficient resolution) can be evaluated on a grid over a hypercube. However, as the number of model parameters increases, the number of required parameter combinations quickly becomes unrealistic (with current computational memory limiting the approach to around six model parameters).

To adequately explain the (2D) data in the divertor, complex diagnostics often necessitate forward models that demand well in excess of six model parameters. Procedures \cite{neal2011mcmc, hoffman2014no} have been developed for inference in such high-dimensional problems. However, such approaches require careful analysis and construction of the posterior distribution (see appendix~\ref{sec:Appendix: compl_posterior}) to ensure the posterior distribution is correctly characterised.


\subsection{Advantages and disadvantages of cell-based versus mesh-based approaches}
\label{sec:cell vs mesh}
To avoid the need for complex analysis procedures and to reduce the computational time required, one can simplify the analysis by restricting the number of model parameters. For example, at TCV, spectrally filtered camera images (at different hydrogen emission wavelengths obtained by MANTIS \cite{Perek2019submitted}) have been tomographically inverted (based on the SART technique \cite{Anderson1984}) to perform inference \cite{Perek_2022}. This transformed the camera data into emissivity ``data-points'' at discrete spatial locations, \textit{cells}, throughout a 2D poloidal cross-section. Such `cell-based' inference partitions the problem into multiple, independent inferences - one for each cell's emissivity ``data-point''. This enables a forward model of the emissivity of a cell rather than a full model of the camera data (which requires accounting for the plasma parameters in all cells simultaneously). Similar cell-based inferences have also been applied to hydrogen Balmer line emission in JET \cite{Karhunen2023} and He I emission in TCV \cite{Linehan2023}.

The cell-based inference comes with some drawbacks:
\begin{enumerate}
    \item Tomographic inversion is typically an ill-posed problem and can result in various inversion artefacts, leading to uncertainties in the emissivity ``data-point'' denoted to each cell.
    \item When an independent inference is made in each cell, information from neighbouring cells is not accounted for. This leads to additional limitations:
    \begin{enumerate}
    \item Cell-based inference reduces the ability to simultaneously include many diagnostics in the overall inference. For example, \cite{Perek_2022} showed, using a cell-based approach, that accurate estimates of neutral densities could be obtained, but \emph{only} at the location of the Thomson scattering measurement (as this strongly constrains \edens{} and \etemp{}). However, since each cell was treated independently, the information was lost outside of the location of the Thomson scattering measurements.
    \item \etemp{}, \edens{} and \ndens{} are not expected to vary wildly within millimetres (a certain spatial smoothness is expected). To include this `prior' information, the inference of one cell would have to influence the inference of another cell and vice versa. This is not possible when each cell is treated independently. 
      \end{enumerate}
\end{enumerate}

Rather than splitting the plasma state into many independent cells, a `mesh-based' inference simultaneously considers plasma parameters at vertices spatially distributed over a mesh. By defining this mesh over a two-dimensional poloidal cross-section, through interpolation between vertices and the assumption of toroidal symmetry, plasma parameter values are available at all spatial locations throughout the divertor. This allows for comprehensive forward models to predict diagnostic data without the need for simplification of the data (e.g. through tomographic inversion). Crucially, such joint treatment of parameters retains information; by demanding a certain level of spatial smoothness of the plasma parameters, discrete information from Thomson scattering measurements influences all nearby vertices. Furthermore, combining diagnostics has a \textit{boot-strapping}\footnote{i.e. information from one diagnostic can improve the information content of another diagnostic and vice versa.} effect on improving inference uncertainties. Consider a camera pixel's chord which passes through a Thomson scattering measurement location, the \etemp{} and \edens{} information provided by the Thomson scattering constrains the emission arising from that region of the plasma which, in turn, limits the emission arising from all other regions of plasma along the chord. The cost of jointly considering plasma parameters is a drastic increase in the total number of model parameters and a more complex posterior distribution. However, previous work \cite{Bowman_2020} demonstrated that the additional information held by not pre-inverting imaging data and jointly considering parameters distributed over a mesh across a poloidal cross-section allowed for inferences with high accuracy despite the additional complexities in the posterior distribution. 

\section{The \divmidas{} Method}
\subsection{A Field-Aligned Inference Mesh}
\label{sec:ida setup} 
As with \cite{Bowman_2020}, we seek a mesh-based inference. Much of the physics in divertor plasmas is tied to magnetic flux surfaces and consequently is spatially-dependent. Bayesian inference permits \emph{a priori} information to our inference in the form of the prior probability of equation~(\ref{eqn:bayes}). Consequently, to enable an efficient manner for including prior information on anisotropic plasma properties, we align our inference mesh to surfaces of constant poloidal magnetic flux. 


\begin{figure}
    \centering
    \includegraphics[width=\linewidth]{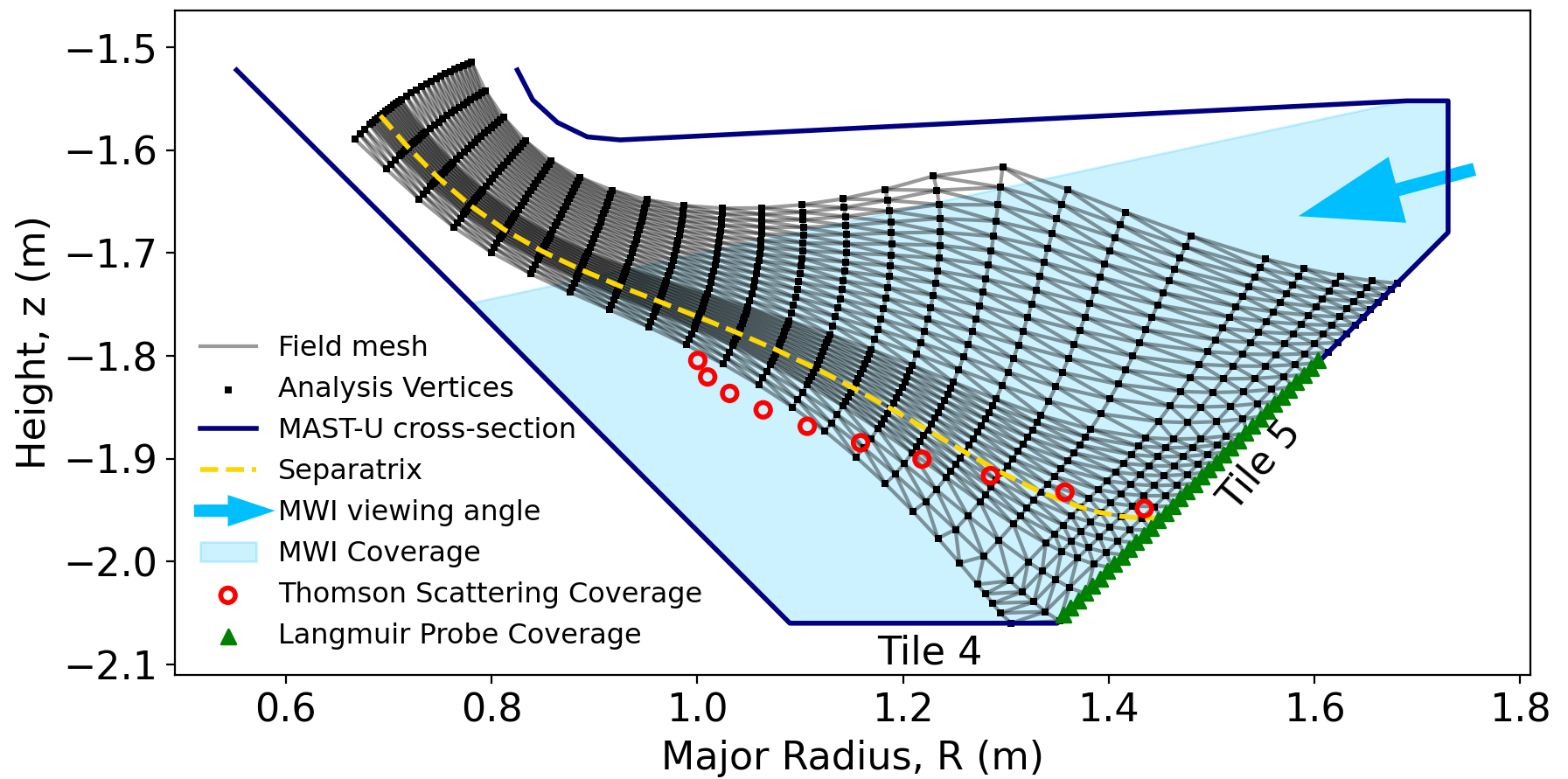}
    \caption{The inference mesh and the diagnostic coverage in the MAST-U divertor. The magnetic geometry is obtained from SOLPS-ITER simulations of the MAST-U Super-X divertor.}
    \label{fig:Diagnostic_Coverage}
\end{figure}

The inference mesh used in this work is shown in Figure~\ref{fig:Diagnostic_Coverage} where $V$ vertices form a two-dimensional grid of normalised poloidal magnetic flux coordinates, $\psi_N$, and distances to the target (Tile 5) parallel to surfaces of constant poloidal magnetic flux. With knowledge of the magnetic pitch angle at each vertex, the magnetic connection length to the target (distance to the target parallel to a flux tube) at each mesh vertex can be trivially recovered. 

\subsection{The Posterior Distribution}
The logarithm of the posterior distribution of equation~(\ref{eqn:bayes}), \logposterior{}, is the sum of the log-likelihood, \loglikelihood{}, log-prior, \logprior{} and a normalisation constant, $c$, which is neglected due to its lack of influence on either the MAP estimate or posterior distribution width (uncertainty). Both the log-likelihood and log-prior are themselves the sum of numerous log-probability distributions, respectively pertaining to different diagnostics, $i$, and different physics-informed priors, $j$. The total log-posterior probability is therefore,
\begin{equation}
    \underbrace{\mathcal{L}\left(\thetavec|\mathcal{D}\right)}_{\mathrm{log-posterior}}
    = c + 
    \underbrace{\sum_i  \mathcal{L}_{i}\left(\mathcal{D}|\thetavec\right)}_{\mathrm{log-likelihood}}
    + 
    \underbrace{\sum_j  \mathcal{L}_{j}\left(\thetavec\right)}_{\mathrm{log-prior}}.
    \label{eqn:log-posterior}
\end{equation}

\subsection{Likelihoods}
For each diagnostic, a multivariate Gaussian likelihood distribution (uncorrelated between diagnostics) was assumed such that each diagnostic log-likelihood (minus an omitted normalisation constant) was given by
\begin{equation}
    \eqalign{
    \mathcal{L}_i\left(\mathcal{D} | \theta\right) =  -\frac{1}{2} \left(\mathbf{d}_{i} - f_{i}\left(\mathbf{\theta}\right)\right)^T\mathbf{\Sigma}_{i}^{-1}\left(\mathbf{d}_{i} - f_{i}\left(\mathbf{\theta}\right)\right).
    }
    \label{eqn:Likelihood}
\end{equation}
$\mathbf{d}_i$ specifies a vector of measurements for the $i^{th}$ diagnostic. $f_i\left(\mathbf{\theta}\right)$ represents the forward model that predicts a diagnostic response from a plasma state (represented by the model parameters, $\theta$). 
$\mathbf{\Sigma}$ is the covariance matrix, which contains information on uncertainties.

For this study, the diagnostic set consisted of:
\begin{itemize}
    \item divertor Thomson scattering (TS) measurements of \etemp{} and \edens{};
    \item  divertor Langmuir probe (LP) measurements of ion saturation current density, $j_{\mathrm{sat}}$, at the target (tile 5);
    \item camera images \footnote{Modelled after MAST-U's Multi-Wavelength-Imaging (MWI) diagnostic \cite{Feng2017,Wijkamp2023} and TCV's MANTIS diagnostic \cite{Perek2019submitted}} of the brightness of 
    \begin{itemize}
        \item Balmer lines \{3($\alpha$),...,7($\epsilon$)\}$\to$ 2,
        \item Helium lines \{502nm, 668nm, 728nm\}.
    \end{itemize} 
\end{itemize}
Thomson scattering measurements of \etemp{} and \edens{}, along with uncertainties, are routinely available at MAST-U \cite{Hawke_2013}. However, data points may be unavailable or not perfectly aligned to specific flux tubes, which we reflect in our chosen positions shown in Figure~\ref{fig:Diagnostic_Coverage}. We assume the availability of Langmuir probe \jsat{} measurements along MAST-U's tile 5 (Figure~\ref{fig:Diagnostic_Coverage}). Assuming similar ion and electron temperatures in the target vicinity, we model
\begin{equation}
   j_{\mathrm{sat}} = \frac{en_{e}}{2}\sqrt{\frac{2eT_{e}}{m_i}}
\end{equation}
where $e$ is the elementary charge and $m_i$ is the ion mass (taken exclusively as deuterium ions). We assume the availability of brightness images (calibrated spatially and in absolute brightness value) for specific wavelength windows associated with each line. To model each line's brightness image, the line's emissivity was modelled at each mesh vertex and a geometry matrix was used to transform
the poloidal cross-section emissivity into camera brightness.

\subsubsection{Emission models}
\label{ch:emissmodel}
The emissivity of a species' atomic transition $m\to n$ (at a spatial location) can be modelled via,
\begin{equation}
\varepsilon^{m\to n} = \sum_i n_{i, 0} n_{i, 1} \pec{}^{m\to n}_{i}(T_e, n_e),
\label{eqn:emissivity}
\end{equation}
where $i$ is a contributing process to the $m\to n$ emission. $n_{i, 0}$ and $n_{i, 1}$ are the number densities of the two species involved in the process. The \etemp{} and \edens{} dependent photon-emissivity coefficient, PEC, of the process is generated from generalised collisional-radiative models (provided by ADAS \cite{adas} for atomic interactions).

\subsubsection{Inclusion of molecular effects in the hydrogen Balmer emission model}
\label{sec:qmol definition}
Hydrogenic Balmer line emission is typically modelled through two predominant atomic processes: electron-impact recombination (EIR) and electron-impact excitation (EIE). However, recent work \cite{Verhaegh_2022, Karhunen2023} has shown that hydrogen molecules can play an important role in both monitored hydrogen emission and plasma dynamics through molecular activated recombination (MAR) and molecular activated dissociation (MAD). MAR and MAD create excited hydrogen atoms and so PMI must be considered as an additional process in equation~(\ref{eqn:emissivity}).

A comprehensive model of PMI contributions to Balmer lines requires knowledge of $D_2$ plasma chemistry. To avoid the need to introduce the various parameters specifying this plasma chemistry, we parameterise molecular contributions to Balmer line emission via a single parameter, \qmol{}, in Appendix~\ref{sec:Appendix: Molecular emission model}. \qmol{} is defined as the ratio of $D_\alpha$'s molecular emissivity contributions, $\varepsilon^{3\to 2}_{\mathrm{mol}}$, to atomic emissivity contributions, $\varepsilon^{3\to 2}_{\mathrm{atm}}$,
\begin{equation}
    \label{eqn:Q_def}
    Q_{D_\alpha}^{\mathrm{mol}} \equiv \frac{\varepsilon^{3\to 2}_{\mathrm{mol}}}{\varepsilon^{3\to 2}_{\mathrm{atm}}} = \frac{\varepsilon^{3\to 2}_{\dtwoeff{}}}{\varepsilon^{3\to 2}_{\mathrm{eir}} + \varepsilon^{3\to 2}_{\mathrm{eie}}}.
\end{equation}
As outlined in Appendix~\ref{sec:Appendix: Molecular emission model}, an effective molecular PEC, $\pec{}^{n\to 2}_{\dtwoeff{}}$, was generated to conglomerate the various molecular contributions to a line's transition. The emissivity of Balmer lines $n\to2$ was thus given by,
\begin{equation}
    \eqalign{
    \varepsilon_{n\to 2} = \underbrace{n_e^2 \pec{}^{n\to 2}_{eir} + n_Hn_e \pec{}^{n\to 2}_{\mathrm{eie}}}_{\mathrm{Atomic\; contribution,}\; \varepsilon^{n\to2}_{\mathrm{atm}}} +\cr\ \underbrace{Q_{D_\alpha}^{\mathrm{mol}}\left(n_e^2 \pec{}^{3\to 2}_{\mathrm{eir}} + n_Hn_e \pec{}^{3\to 2}_{\mathrm{eie}}\right)\frac{\pec{}_{\dtwoeff{}}^{n\to 2}}{\pec{}_{\dtwoeff{}}^{3\to 2}}}_{\mathrm{Molecular\; contribution,}\; \varepsilon^{n\to2}_{\mathrm{mol}}} .}
    \label{eqn:Balmer_emissivity}
\end{equation}
Consequently, no molecular density was explicitly required. Furthermore, $\pec{}_{\dtwoeff{}}$ was only present as a ratio between lines $n\to 2$ and $3\to 2$ minimising the impact of the approximations required for $\pec{}^{n\to 2}_{\dtwoeff{}}$. This allowed the inclusion of molecular processes by a single field parameter with approximations that had minimal impact.

\subsubsection{Helium emission model}
MAST-U's MWI and TCV's MANTIS cameras also routinely capture Helium-I emission lines which have been shown \cite{Linehan2023} to provide information on \edens{} and \etemp{}. To avoid inaccuracies in Helium camera forward models, we explicitly include parameters for the densities of neutral and singly charged helium (\hendens{} and \hepdens{} respectively) at each mesh vertex (at a cost of 2V additional free parameters) in our inference. This allowed both EIE and EIR contributions to be modelled via equation~(\ref{eqn:emissivity}). Following the work of \cite{Biggelaar2022}, 668, 728 and 502 nm singlet He I lines were chosen due to their insensitivity to the transport of metastable states. Additional details and potential complexities are outlined in Appendix~\ref{sec:Appendix: Helium details}.

\subsection{Priors}
We refer to each physics-based constraint considered in the inference as a prior. Each prior is imposed as a contribution, $\mathcal{L}_j\left(\theta\right)$, to the overall log-prior of equation~(\ref{eqn:log-posterior}) with details of the probability function used provided in Appendix~\ref{sec:Appendix: Prior Distributions}. Each prior is evaluated at all mesh vertices relevant to that prior.


\subsubsection{Spatially-Independent and -Dependent Priors}
We consider `spatially-independent priors' to be those that, when being evaluated at a certain mesh vertex, require no information about parameter values at a different mesh vertex. Consequently, these can be included in both the cell-based and mesh-based inference paradigms. They include priors that consider the interplay between parameters at a single mesh vertex such as an upper bound of the static electron pressure as well as bounds on parameter values. All spatially-independent priors used in this work are detailed in Table~\ref{tab:standard priors} of Appendix~\ref{sec:Appendix: Prior details}.

`Spatially-dependent priors', when being evaluated at a certain mesh vertex, require information of parameter values at other mesh vertices and so cannot be used in the cell-based inference paradigm. An example of such a prior is the expected monotonic reduction (toward the target) in the heat flux parallel to a surface of constant poloidal magnetic flux. All spatially-dependent priors used in this work are detailed in Table~\ref{tab:grid-enabled priors} of Appendix~\ref{sec:Appendix: Prior details}.

\subsection{Inference Parameters}
In the mesh-based inference, the model parameters,
\begin{equation}
    \theta = \left\{F_1^{(1)},...,F_1^{(V)}, ..., F_N^{(1)},...,F_N^{(V)}\right\},
    \label{eqn:parameter_vector}
\end{equation}
specify the value of each of $N$ fields, $F$, at each of $V$ mesh vertices.
To compute forward models for all diagnostics, we require the parameter field set
\begin{equation}
    F = \left\{T_e, n_e, n_H, Q_{\mathrm{mol.}}, n_{He^0}, n_{He^{1+}}\right\}.
\end{equation}

\etemp{}, \edens{} and \ndens{} are of direct interest for divertor physics. These, as well as \qmol{}, can be used to derive further quantities of interest to divertor physics including: static electron pressure; heat flux parallel to flux tubes; divertor ionisation sources; ion sinks (volumetric recombination due to electron-ion recombination and molecular activated recombination); hydrogenic power losses; and more \cite{Verhaegh2021a}. \hendens{} and \hepdens{} parameters can be identified as nuisance parameters which are necessary to evaluate forward models but not of direct interest in this work.
\section{Synthetic Data Setup}
\label{sec:Synthetic Data}
\begin{figure*}
    \centering
    \includegraphics[width=\linewidth, valign=t]{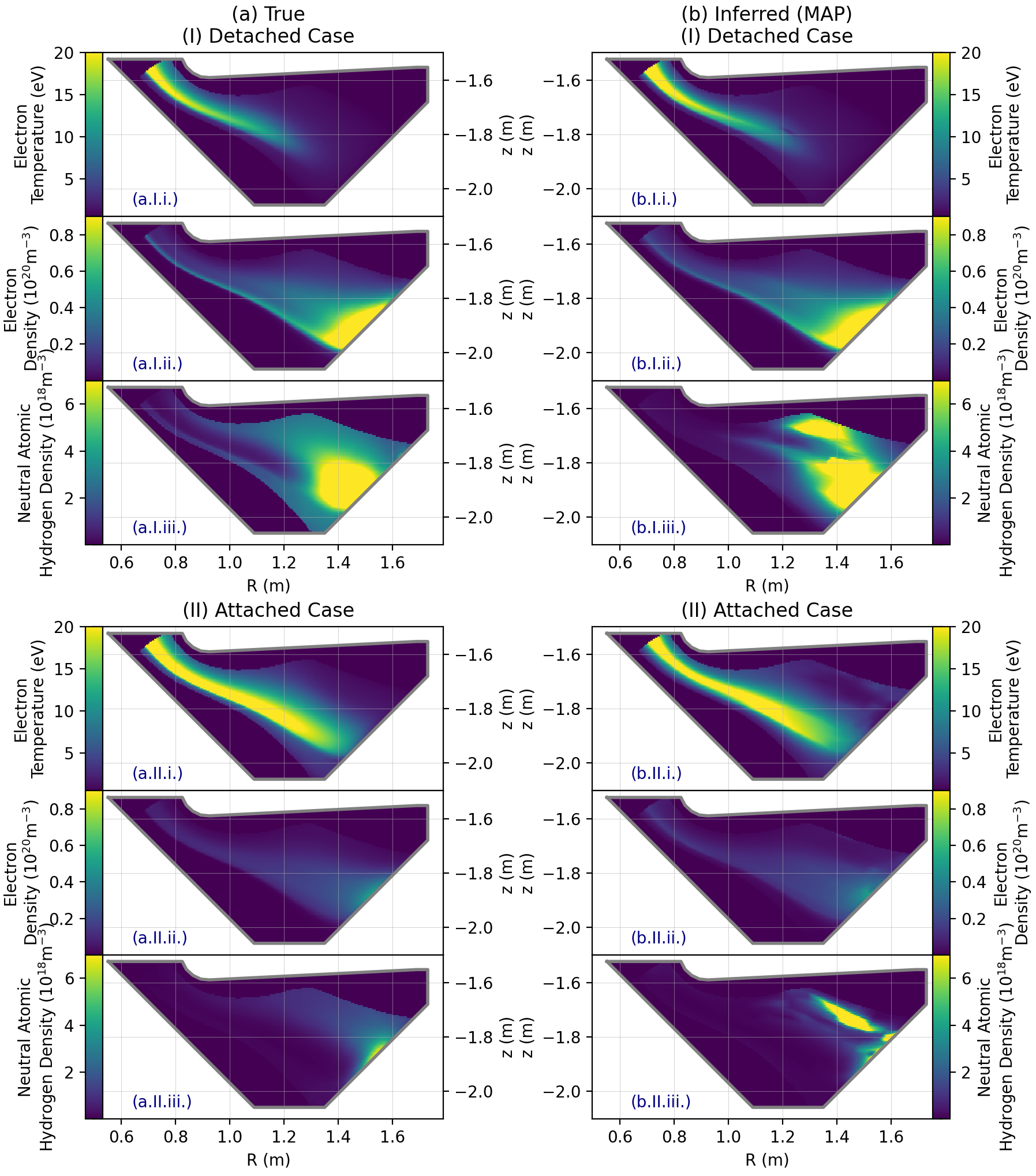}
    \captionof{figure}{
    Comparison of `true' fields used to generate synthetic data (a) and the inferred MAP estimate (b).
    \newline
    (a)
    Electron temperature and density fields for two synthetic cases from SOLPS-ITER. 
    \newline
    \qquad (I)\; The \state{1} follows a detached scenario where the \etemp{} (a.I.i.) drops below 5eV before it reaches the target. 
    \newline
    \qquad (II) The \state{2} follows an attached scenario in which the \edens{} (a.II.ii.) peaks at the target.
    \newline
    (b) 
    Maximum a posteriori (MAP) estimate inference of scenarios displayed in Figure \ref{fig:Results}(a) using posterior \posterior{1}. MAP was found using the generational algorithm outlined in appendix A of~\cite{Bowman_2020}.}
    \label{fig:Results}
\end{figure*}

The new integrated data analysis methodology developed in this work was applied to synthetic data. The data were generated from synthetic diagnostics applied to two SOLPS-ITER simulations: a `\state{1}', shown in Figure~\ref{fig:Results}aI, where the plasma was in a detached state in which PMI was expected to play a significant role; and an `\state{2}', shown in Figure~\ref{fig:Results}bII, where the plasma was in an attached state in which PMI was not expected to play a significant role. These simulations had an input power of 2.5 MW and were obtained from \cite{Fil_2022} using the `Super-X low $\alpha$' geometry \footnote{This geometry is observed more easily by the synthetic imaging diagnostic as the Super-X has no curvature in the separatrix near the target, see Figure~2 of \cite{Fil_2022}.}. A trace amount of helium was added to get realistic relative 2D spatial distributions of \hendens{} and its ratio to \hepdens{}, although the absolute numbers for \hendens{} may not be characteristic of that obtained in the tokamak \footnote{However, the absolute magnitude of \hendens{} is not expected to impact our inference.}. 

For each case, the SOLPS-ITER simulation was taken as the `true' plasma state from which the relevant fields were extracted. To compute \qmol{}, $\mathrm{PEC}_{\mathrm{D{2,eff}}}$ was used to post-process the $\mathrm{D}_\mathrm{2}$ density into the densities for the various molecular ions. It should be noted that this leads to significantly higher $\mathrm{D}_\mathrm{2}^+/\mathrm{D}_\mathrm{2}$ ratios than reported by SOLPS-ITER directly, which is known to under-predict the $\mathrm{D}_\mathrm{2}^+/\mathrm{D}_\mathrm{2}$ ratio due to inaccuracies in the molecular charge exchange rate (see \cite{Verhaegh2021,Verhaegh2023c,Verhaegh2023a}). 

To create the synthetic data sets, for each case, the forward models of equation (\ref{eqn:Likelihood}) were used to predict diagnostic response and Gaussian errors akin to those expected in experiments were added. As a comparison metric, for each field, the median absolute percentage error (MdAPE) between the `true' plasma state and the inferred MAP estimate at each mesh vertex was found.

\section{Results}
\label{sec:Posterior Distributions}
We have applied the \divmidas{} framework to the two synthetic data sets outlined in Section~\ref{sec:Synthetic Data} (with both including PMI in the measured Balmer-line brightness). Four different posterior distributions have been constructed under the framework to ascertain the influence of the advanced emission model, the use of spatially-dependent priors, and the inclusion of additional diagnostics. These posterior distributions are:

\begin{itemize}
    \item [\posterior{1}] 
        Likelihood: TS, LP, helium and hydrogen Balmer multi-wavelength imaging diagnostics with a forward model that includes PMI contributions to Balmer emission
        
        Prior: spatially-dependent and spatially-independent priors

        This is the most complete version of the \divmidas{}.
    \item [\posterior{2}] 
        As with \posterior{1}, but with a forward model that doesn't include PMI contributions to Balmer emission (\qmol{}=0)
    \item [\posterior{3}] 
        As with \posterior{1}, but without spatially-dependent priors
    \item [\posterior{4}] 
        As with \posterior{1}, but without helium multi-wavelength imaging
\end{itemize}

\subsection{Synthetic diagnostic test results of the full \divmidas{} analysis}
\label{sec:fullIDA-result}
\begin{figure} 
    \centering
    \includegraphics[width=\linewidth]{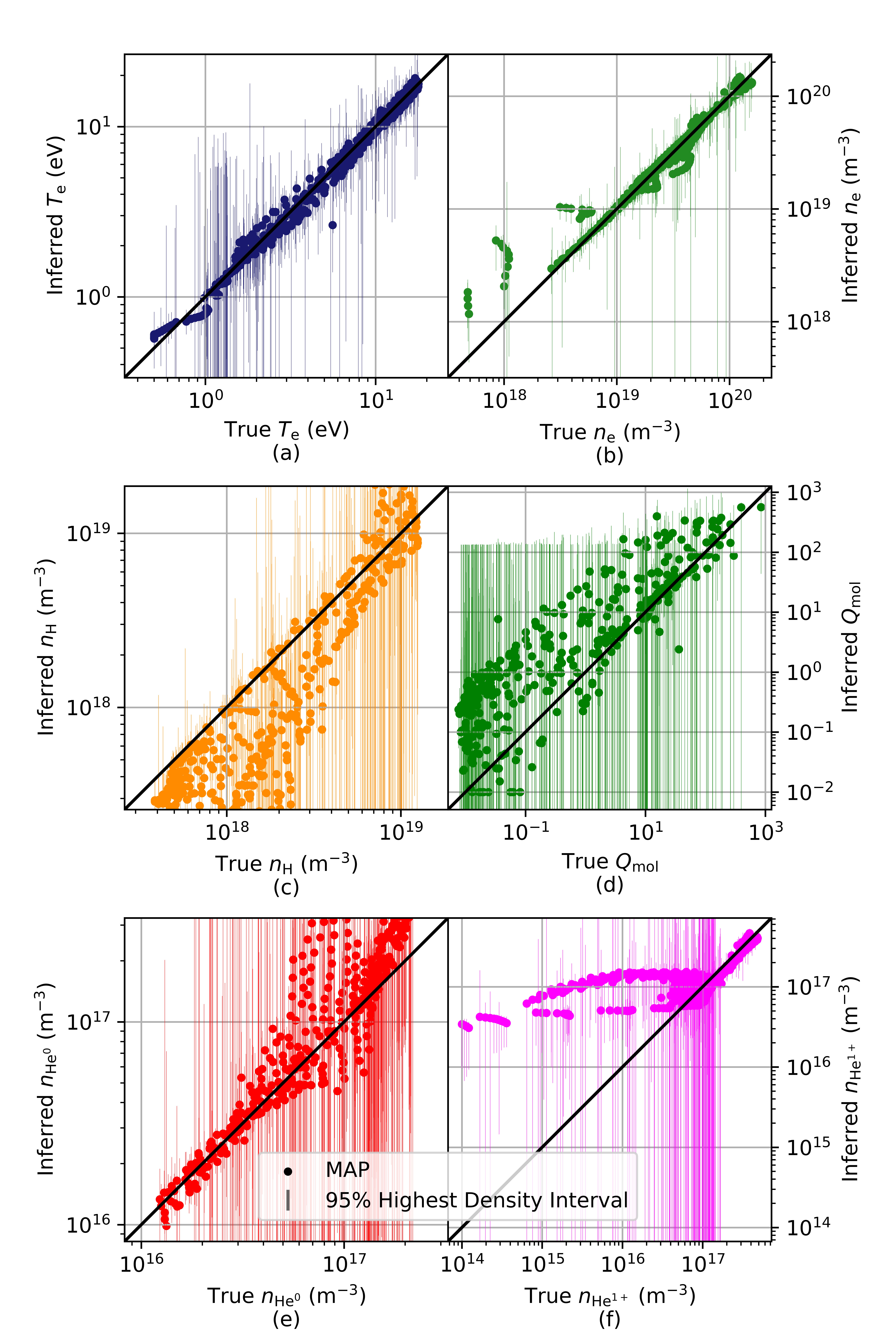}
    \caption{Comparison of true and inferred (MAP) parameter values using posterior \posterior{1} for the \state{1}. The 95\% highest density interval (HDI) for the inferred parameters are displayed. The percentage of true data points lying within the 95\% HDI are 95\%, 88\%, 63\%, 91\%, 95\%, 63\% for the \etemp{}, \edens{}, \ndens{}, \qmol{}, \hendens{} and \hepdens{} respectively.
    }
    \label{fig:MEX Scatter}
\end{figure}

As shown in Figure~\ref{fig:Results}, the most complete version of the \divmidas{} framework (posterior \posterior{1}) was able to correctly infer the general spatial distribution of the electron temperature and electron density in both the \state{1} and the \state{2}. The atomic neutral density was less successfully inferred; however, the most prominent discrepancies occurred in regions of low Balmer and He I emission (deep into the common flux region, away from the separatrix). The inference achieved an MdAPE of 6\%, 5\% and 33\% for \etemp{}, \edens{} and \ndens{} respectively for the \state{1} and 6\%, 3\% and 9\% for the \state{2}.

The inferred uncertainties in the parameters (their 95\% HDI) were found to be largely in agreement with the error between the true parameter values and the inferred MAP estimates. This is demonstrated in Figure~\ref{fig:MEX Scatter}. The uncertainty was found to be substantially greater for \ndens{} and \qmol{} parameters, (as well as the nuisance parameters \hendens{} and \hepdens{}) compared to \etemp{} and \edens{} parameters. \etemp{} and \edens{} parameters with large uncertainty were found to be in regions out of view of the synthetic camera diagnostic (upstream) and in regions of low Balmer and He I emission (deep into both the private flux and common flux regions).  Figure~\ref{fig:uncertainty_quantification} of Appendix~\ref{sec:Appendix: compl_posterior} further demonstrates the inferred uncertainty was able to reflect the error in the inferred parameters.

\subsection{Performance of the different posterior distributions}
\label{sec:Posterior Distribution Comparisons}

\begin{figure*}
    \centering
    \includegraphics[width=.85\linewidth, valign=t]{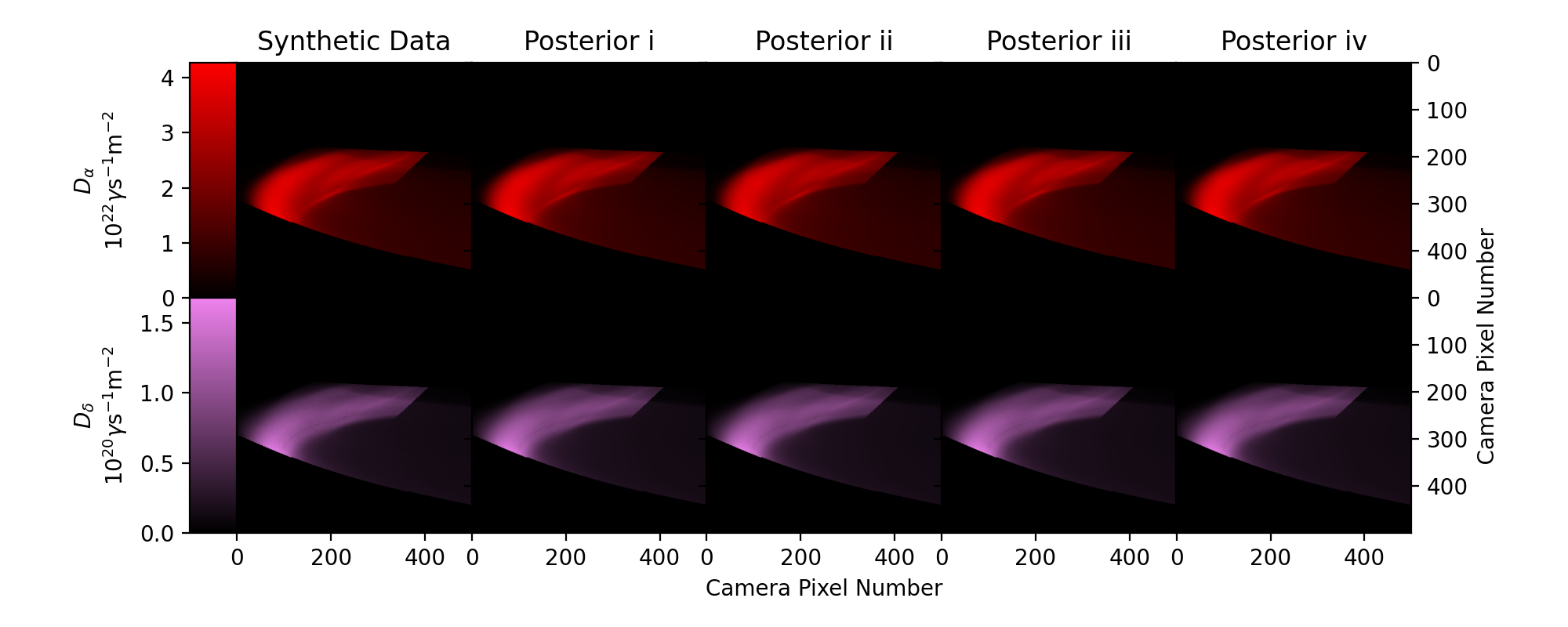}
    \captionof{figure}{
    The synthetic camera data for Balmer lines $D_\alpha$ (3$\to$2, red) and $D_\delta$ (6$\to$2, lilac) and their forward model predictions based on the MAP estimate of the four posterior distributions outlined in Section~\ref{sec:Posterior Distributions}. 
    }
    \label{fig:Cameras}
\end{figure*}
 
\begin{figure}
    \centering
    \includegraphics[width=\linewidth]{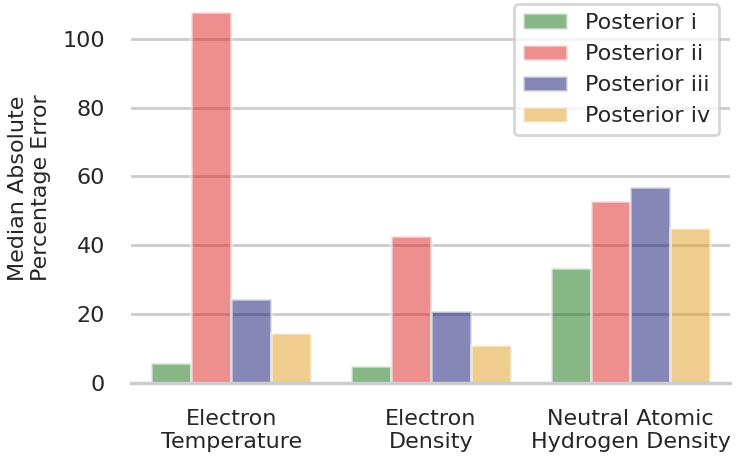}
    \caption{Performance of posteriors \posterior{1}-\posterior{4} across fields of interest for the \state{1}.}
    \label{fig:Percentage_Comparison}
\end{figure}

Inference with each posterior distribution was able to accurately recreate the synthetic data, as highlighted in Figure~\ref{fig:Cameras}. However, the inferred parameters (MAP estimate) of each posterior distribution were considerably different. The resultant MdAPE scores of inference with each of the posterior distributions outlined in Section~\ref{sec:Posterior Distributions} are shown in Figure~\ref{fig:Percentage_Comparison}.  All parameters of interest were inferred most accurately by the most complete version of \divmidas{} (posterior \posterior{1}). We will now discuss each reduced analysis case (posteriors \posterior{2}-\posterior{4}) in more detail.

\subsubsection{Importance of including PMI in the Balmer emission model}
\label{sec:Importance of the molecular model}
Posterior \posterior{2} only considers atomic contributions to Balmer line emission which results in a sustained overestimation of the electron temperature outside of the region where electron-ion recombination emission is dominant as can be seen 
in Figure~\ref{fig:Te at psi=1.0}. The cause of this was that the Balmer line emission is increased by the presence of plasma-molecule interaction. Because the Balmer line ratios from PMI and EIE are similar and significantly different from EIR, these brighter images could not be explained by EIR. Therefore, the model omitting PMI was compelled to attribute these bright images to EIE which, in turn, demanded higher electron temperatures, leading to the overestimate. This overestimation is an important demonstration that an incorrect model in Bayesian inference does not necessitate the true solution to fall within the found uncertainty interval if an invalid emission model is applied. This highlights the importance of using a model that accommodates PMI when inferring divertor plasmas with substantial molecular hydrogen densities.

\subsubsection{Importance of Spatially-Dependent Priors}
\label{sec:Importance of Spatially-Dependent Priors}
The spatially-dependent priors are permitted through jointly conducting inference with parameters at each mesh vertex as allowed by the mesh-based inference paradigm. Figure \ref{fig:ne at psi=1.0} demonstrates that the omission of these in posterior \posterior{3} substantially increased uncertainty on the electron density (which was also observed for other inferred parameters) and the complete loss of information out of view of the synthetic camera diagnostic (as expected). The lack of additional information on trends in static electron pressure and smoothness leaves an ill-posed problem in which many different combinations of parameters can accurately recreate the multi-wavelength imaging data, thus causing large uncertainty in the inference. This highlights the benefit of jointly considering many parameters at different spatial positions in the inference as permitted by the mesh-based \divmidas{}.

\subsubsection{Importance of helium emission}
\label{sec:Importance of Helium}
Posterior \posterior{4} reflects situations where helium data are unavailable or unreliable. The helium singlet lines are known to carry information on both the electron temperature and electron density. The exclusion of helium data degraded the MdAPE for \etemp{} and \edens{} to 14\% and 11\% respectively, as shown in Figure \ref{fig:Percentage_Comparison}. 

\begin{figure}[!h]
    \centering
     \begin{subfigure}[b]{\linewidth}
        \centering
        \includegraphics[width=1.1\linewidth]{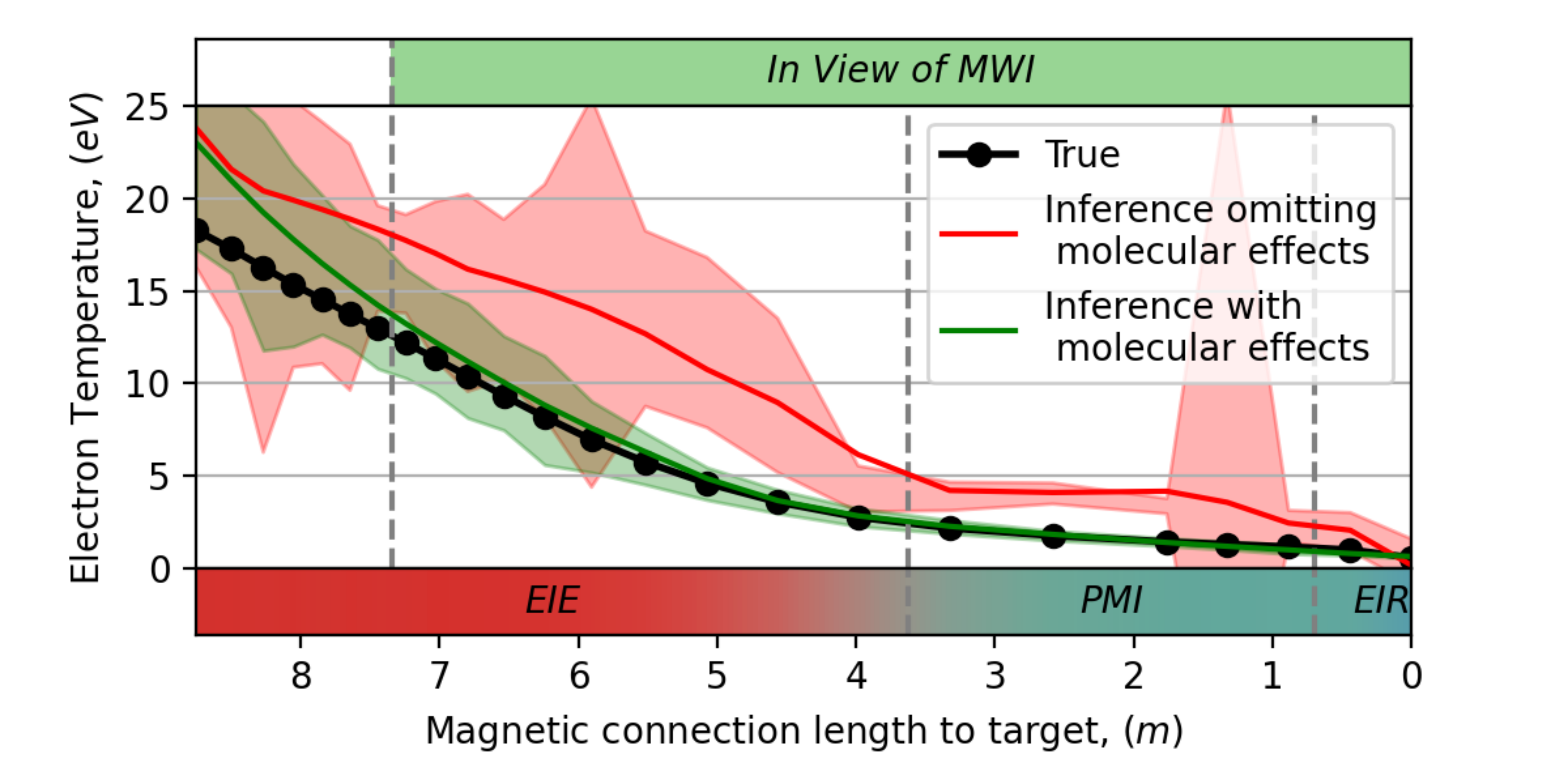}
        \caption{\etemp{}\, comparison of posterior \posterior{1} (green) and posterior \posterior{2} (red).}
        \label{fig:Te at psi=1.0}
     \end{subfigure}
     \vfill
     \begin{subfigure}[b]{\linewidth}
        \centering
        \includegraphics[width=1.1\linewidth]{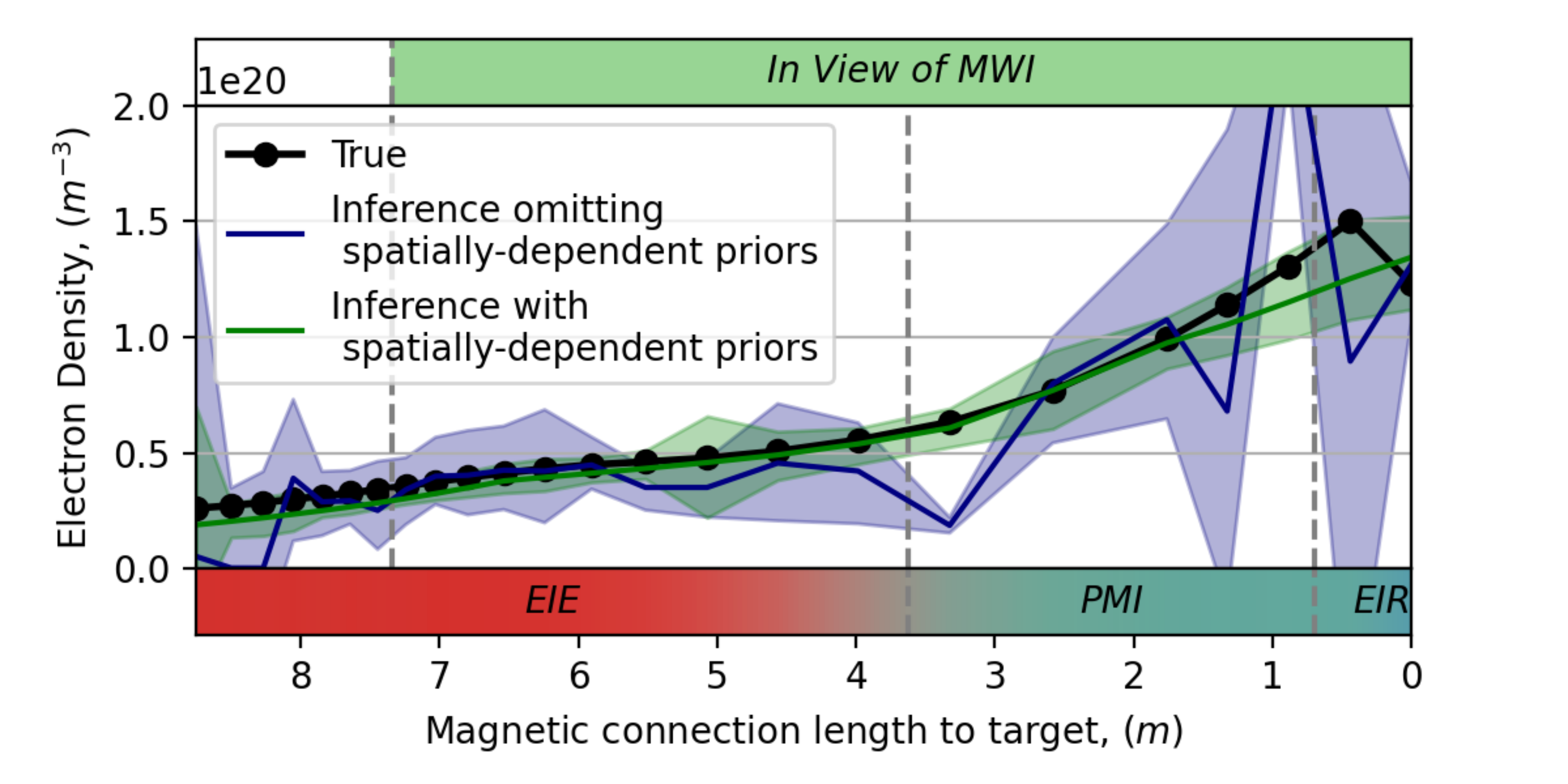}
        \caption{\edens{}\, comparison of posterior \posterior{1} (green) and posterior \posterior{3} (blue).}
        \label{fig:ne at psi=1.0}
     \end{subfigure}
     \begin{subfigure}[b]{\linewidth}
        \centering
        \includegraphics[width=1.1\linewidth]{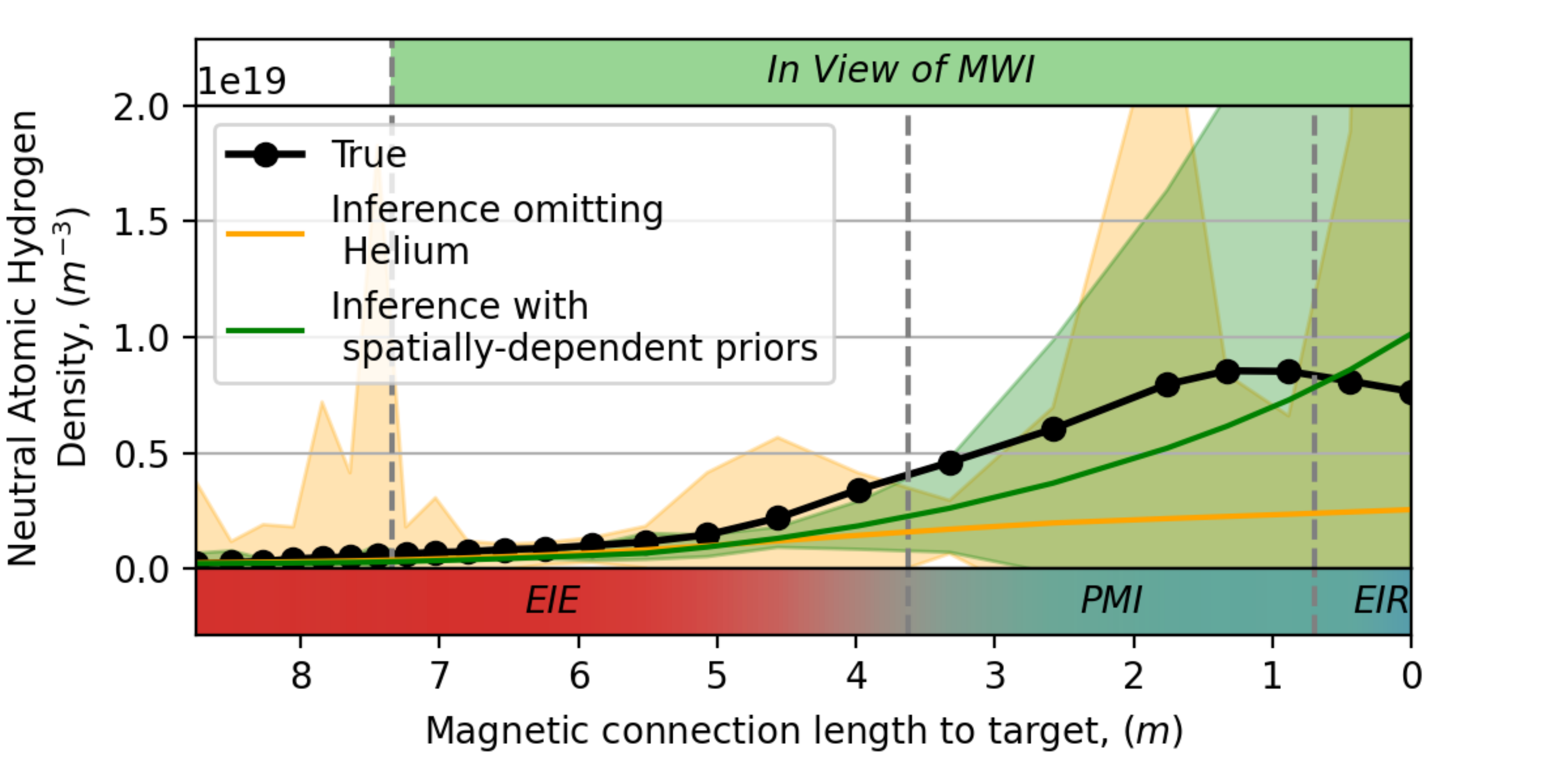}
        \caption{Neutral atom density comparison of posterior \posterior{1} (green) and posterior \posterior{4} (gold).}
        \label{fig:n0 at psi=1.0}
     \end{subfigure}
    \caption{Comparison of inferred parameters with the `true' values (Figure~\ref{fig:Results}Ia) for the \state{1}. The parameters follow a surface of constant poloidal flux, $\psi_N = 1$ (the separatrix). Solid lines indicate the inferred (MAP) parameters and the shaded region indicate the 95\% HDI.}
    \label{fig:X at psi=1.0}
\end{figure}

Along the separatrix, the inferred MAP estimate of posterior \posterior{4} was in good agreement with the true plasma values; however, the inclusion of helium data in posterior \posterior{1} reduced the uncertainty in the inference of \etemp{} and \edens{} (Figure~\ref{fig:Percentage_Comparison}). As shown in Figure~\ref{fig:n0 at psi=1.0}, the inclusion of helium data was found to significantly improve the \ndens{} inference along the separatrix despite the \ndens{} parameter not appearing in the helium forward models. This was because the reduced uncertainty in the inference of \etemp{} and \edens{} managed to better constrain the process leading to Balmer-line emission (EIR/PMI/EIE). By having a good understanding of how much EIE emission (which is \ndens{} dependent) occurs, the \ndens{} inference improved. This serves to demonstrate how the inclusion of additional diagnostics in the IDA can improve the accuracy of the inference of certain parameters even if they themselves are not involved in the forward model of the additional diagnostic.

\section{Discussion}

In this section, we will discuss the limitations of the current analysis and possible improvements for the future with an emphasis on the preparation of applying \divmidas{} to experimental data. We will go on to discuss possible applications of this analysis and what the implications are for such techniques.

\subsection{Limitations of the current \divmidas{} and possible improvements}
\subsubsection{Consideration for experimental data application: forward model inaccuracies}

As observed in section~\ref{sec:Importance of the molecular model}, inaccurate forward models can lead to MAP estimates with an inferred uncertainty range that is inconsistent with the `true', known result despite achieving close agreement to the data (seen in Figure~\ref{fig:Cameras}). There are uncertainties in the diagnostic descriptions that will cause inaccurate forward models. Such uncertainties are not observed when performing synthetic tests (since the forward models used for inference are the same as those used to generate synthetic data). Keeping the possibility of such uncertainties in mind is particularly important for experimental data application.

The diagnostic description of imaging and spectroscopic data includes multiple uncertainties. Multi-wavelength imaging employs bandpass filters to monitor a small, spectrally integrated part of the emission spectra. Plasma background emission and neighbouring spectral transitions can contaminate the monitored signal, resulting in overestimated camera brightnesses \cite{Perek2021}. Reflections can displace the monitored emission which, if present, would require a more complex diagnostic description \cite{Karhunen2023}. Similarly, photon opacity \cite{Pshenov2023, Lomanowski2020, Lawson_2024}, if present, would require a substantial modification to the diagnostic description. Additionally, there are uncertainties in the spatial and absolute calibrations of imaging and spectroscopy systems. The various photon emission coefficients used in the emission models can also have significant uncertainties of 10-20\% \cite{Verhaegh2021b}.

The additional uncertainties outlined above require that, when applied to experimental data, the forward model uncertainties (held in $\Sigma$ of equation~(\ref{eqn:Likelihood})) will exceed those from the diagnostic measurements alone. In turn, the uncertainties on inferred parameters will increase from those observed in Figure~\ref{fig:X at psi=1.0}. To limit the impact of the increase in forward model uncertainties, there are several methods that can be used:

\begin{itemize}
    \item Physics and diagnostic studies to ascertain the most appropriate forward models to use, for example using an emission model that supports PMI (as introduced in section~\ref{sec:qmol definition}).   
    \item Introduction of additional model (nuisance) parameters describing uncertainties, for instance for the absolute calibration.
    \item Monte Carlo approaches can be used to perturb uncertain quantities within reasonable intervals (e.g., perturb spatial camera calibration uncertainties \cite{Perek_2022}) to create multiple posterior distributions that can be combined.
    \item Investigating the extent of forward model uncertainties via synthetic testing with intentional mismatches between the forward model used to generate synthetic data and the model used for inference (as performed in section~\ref{sec:Importance of the molecular model}).
      \end{itemize}

\divmidas{} provides a framework for comparing the consistency between the data obtained by the different diagnostics. This makes integrated data analysis particularly suitable for investigating errors in diagnostic descriptions (forward models). This can be strengthened by including additional diagnostics in the \divmidas{} inference (Section~\ref{ch:additional_diags}). Awareness of inadequate diagnostic descriptions can be used to improve instrument characterisation \cite{Perek2019submitted,Perek_2022} (e.g., improved calibrations) and to inform where more comprehensive models are required (e.g., modelling directly to the camera image data instead of pre-inverted data \cite{Bowman_2020}).

\subsubsection{Improving \divmidas{} analysis through additional diagnostics}
\label{ch:additional_diags}

The inclusion of additional diagnostics has multiple benefits. They can: reduce parameter inference uncertainty and improve inference accuracy; mitigate model uncertainties; expose deficiencies in diagnostic models; and act to validate diagnostic interpretation. 
However, care must be taken that the inclusion of an additional diagnostic does not result in too large an increase in the number of free parameters required. Examples of additional diagnostics that can be included in the current \divmidas{} setup without requiring additional free parameters include:

\begin{itemize}
    \item Line-of-sight spectroscopy in the UV-Visible and VUV regime can monitor different hydrogen and helium emission lines which can be modelled using the current parameter set.
    \item Bolometry (including higher resolution imaging bolometry \cite{Federici2023}) can provide additional information on the total radiation. This provides an upper-limit to the permitted hydrogenic radiation (of which can be modelled from the current parameter set).
    \item IR cameras provide target heat flux profiles. This provides an upper-limit on the heat flux calculated to be flowing from the core plasma which can be modelled from the current parameter set as described in section~\ref{sec:derived-quantities}.
\end{itemize}  


\subsection{Implication and relevance of \divmidas{} for furthering exhaust understanding}

The divertor region is complex to diagnose for multiple reasons:
\begin{enumerate}
    \item Its nature is inherently 2D/3D, therefore one cannot map the obtained results to a 1D profile assuming parameters are constant at each flux tube (as is commonly done when studying core plasma profiles).
    \item The number of species that determine the divertor physics is very large - it cannot be approximated as a fully ionised plasma: neutral atoms, neutral molecules, hydrogen ions and impurity ions at a large range of different charge states all impact the divertor.
    \item The divertor state depends on plasma-neutral interactions, which are not field-aligned. Additionally, understanding the divertor state requires detailed knowledge of the plasma chemistry, the different (meta)stable excited levels of neutrals (including vibrationally excited molecules and impurity meta-stable states).
    \item A wide range of different parameters and processes occur in the divertor, many of which are not fully understood but have an impact on the divertor state and diagnostic interpretation. 
\end{enumerate}

The divertor region complexities require the use of information from a wide range of diagnostics. The implementation of \divmidas{} in this work makes advances in all four of these areas:

\begin{enumerate}
    \item \divmidas{} combines the information of multiple diagnostic measurements in different locations allowing inference of 2D profiles of the divertor parameters using a robust statistical framework.
    \item \divmidas{} provides inferences of both electron and ion populations (\edens{}, \etemp{}, \hepdens{}) as well as neutral populations (\ndens{}, \hendens{}, \qmol{} - which provides some indirect information on the molecular content).
    \item \divmidas{} accounts for the complex plasma-neutral interactions (atomic and molecular) that impact the various emission diagnostic measurements. \divmidas{} can handle the non-field-aligned nature of the neutrals by enforcing isotropic smoothness (rather than field-aligned smoothness as with the electron and ion populations).
    \item Combining information from different diagnostic measurements rigorously assesses the consistency between the different diagnostics. \divmidas{} offers a mechanism to investigate the impact of different diagnostic forward models. The parameters inferred by \divmidas{} can be used to derive other quantities that give insight into plasma processes (Section~\ref{sec:derived-quantities}).
\end{enumerate}

The application of this technique to MAST-U data is planned in the future. With these analysis advancements, \divmidas{} can be used to infer 2D profiles of both electron and neutral plasma parameters. This would provide additional information on the divertor state, which can be used to improve the understanding of tokamak divertors, both in conventional divertor designs and alternative divertor designs. 

\subsubsection{Derived Quantities}
\label{sec:derived-quantities}
\begin{figure}
    \centering
    \includegraphics[width=\linewidth, valign=t]{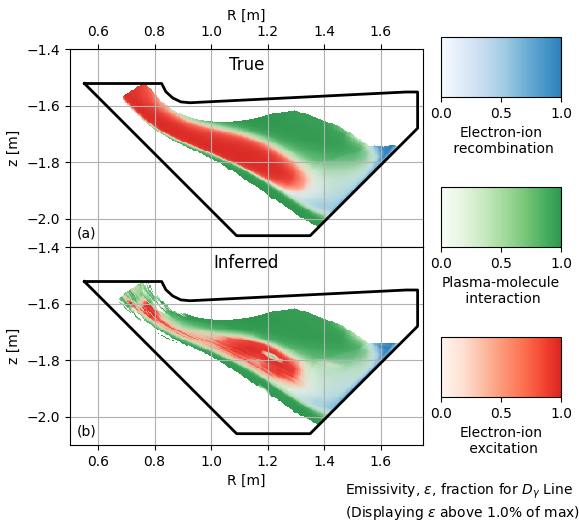}
    \caption{Comparison of true (a) and inferred (b) (using posterior \posterior{1}) dominant emission regions for the D$_\gamma$ Balmer line for the \state{1}.}
    \label{fig:EmissivityFractions}
\end{figure}

\begin{figure}
    \centering
    \includegraphics[width=\linewidth, valign=t]{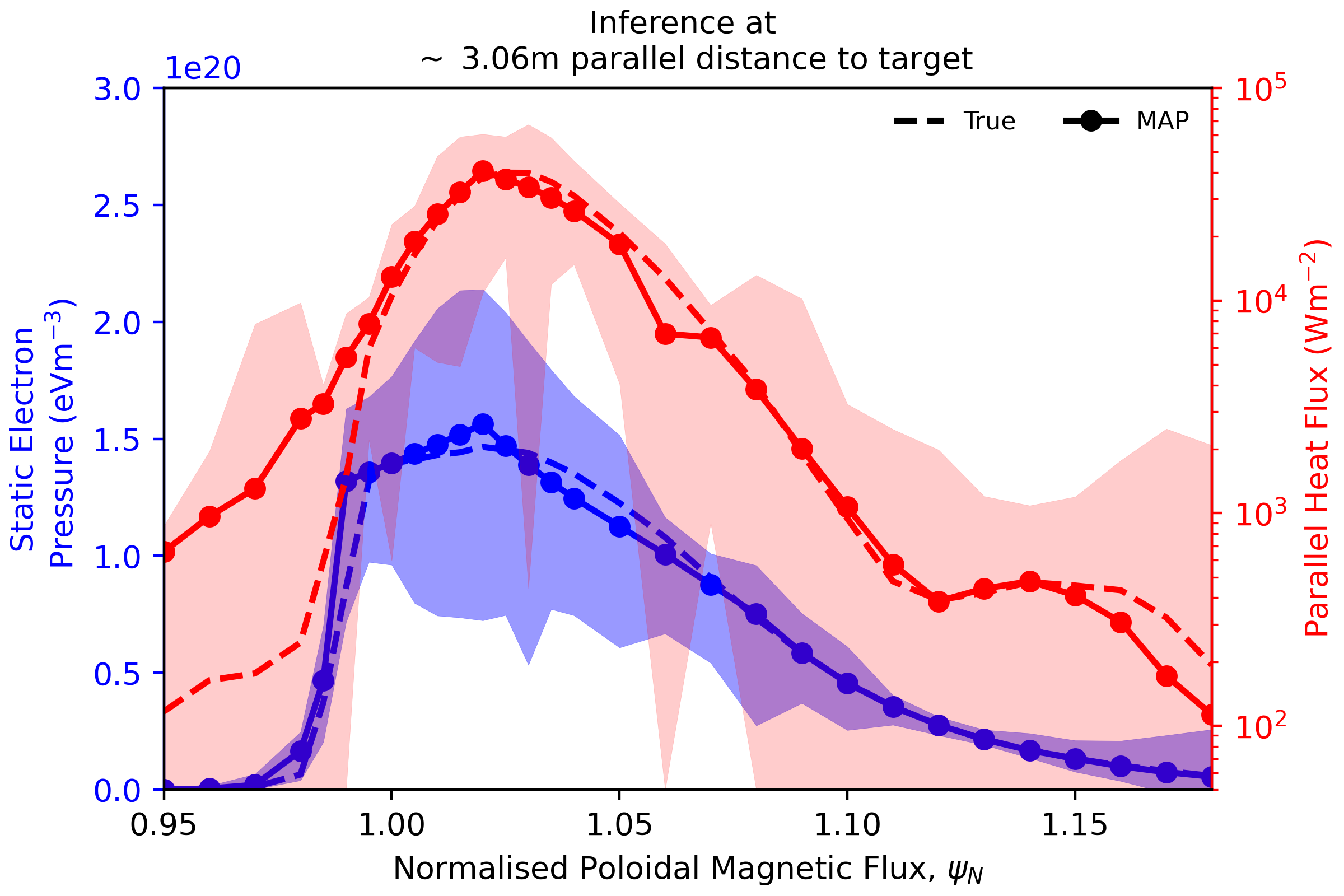}
    \caption{Comparison of true and inferred (using posterior \posterior{1}) static electron pressure and parallel heat flux values for the detached case. Evaluated for parameters perpendicular to surfaces of constant poloidal magnetic flux at a magnetic connection length of around 3 metres to the target}
    \label{fig:DerivedParameters}
\end{figure}

Further derived parameters can be obtained from electron and neutral parameter information. The multi-dimensional posterior distribution provides correlated samples which can be used to compute derived parameters, including their uncertainty. The inferred parameter set can be used to obtain: static electron pressure \cite{Lipschultz2007a}; heat and ion fluxes parallel to flux tubes \cite{Verhaegh2021b}; estimates of ion sources and sinks (both due to atomic (EIE, EIR) as well as molecular processes (MAR, MAI)) \cite{Verhaegh2021a} and (hydrogenic) power loss estimates \cite{Verhaegh2023b,Verhaegh2021a}. This technique was utilised to compute the EIR, EIE and PMI emission fractions for $D\gamma$ in Figure~\ref{fig:EmissivityFractions}. In Figure~\ref{fig:DerivedParameters}, the derived parallel heat flux and static electron pressure are shown at a roughly constant magnetic connection length to the target in the vicinity of the \state{1}'s detachment front. Their associated uncertainties are considerable, however, due to the compounding of uncertainties from multiple parameters.

\subsection{Relevance of \divmidas{} for reactors}
The measurements obtained by \divmidas{} can be used to support comparisons between experiments and plasma-edge simulations to an unprecedented level of detail. Such validation exercises can improve our ability to model divertor plasmas. This is crucial for reducing uncertainties when such models are used to extrapolate current knowledge to reactor-class devices.

Reactors will have to operate with a reduced set of diagnostics due to the harsher reactor conditions (e.g. a neutron-rich environment, intense heat fluxes, limitation to robotic access) and port space will be limited \cite{Wenninger2014}. Reactor diagnostics may fail and become redundant during operation. \divmidas{} provides a robust statistical framework for investigating how much information a set of diagnostics can provide. This can be used to find the diagnostic coverage required to produce a reactor's acceptable level of information on the divertor plasma state and the impact of losing a diagnostic from that set. Langmuir probes measure ion fluxes to chamber walls which are widely used in the community to quickly interpret the plasma state and to understand the power being received by the chamber walls. However, Langmuir probes are unlikely to be operational in harsh reactor conditions. So long as inferences can still be made with sufficient accuracy, the \divmidas{} framework can still provide such quantities even in the absence of the diagnostic.

In reactors, real-time control of the power exhaust problem is required. This requires understanding what information from the divertor (for example, detachment front position) will need to be controlled in real-time. The insight provided by applying \divmidas{} to experimental data can be used to answer that question. Although \divmidas{} is too numerically intensive to be performed in real-time currently, we envisage it to be an important aid (in training, validation, and uncertainty quantification) for neural-network approaches to predicting the divertor state.

Reactor conditions will be significantly different from the MAST-U Super-X plasma conditions and synthetic diagnostics used to test the current \divmidas{} methodology. This will likely require modifications to our forward models, such as the consideration of photon opacity in our emission models \cite{Pshenov2023}. 
Although such \divmidas{} analysis tools hold a strong potential for reactors, substantial effort will be required to test such approaches in reactor-relevant divertor conditions using a reactor-relevant diagnostic set. However, as outlined in this work, the insight into \divmidas{}'s abilities and limitations can be gleaned through synthetic data without needing to wait for reactors to begin operation.

\section{Summary and conclusions}
This study outlines a comprehensive approach for utilising multiple diagnostics to infer two-dimensional fields of plasma parameters, including electron temperature (\etemp{}), electron density (\edens{}), and neutral density (\ndens{}), across the divertor. The developed \divmidas{} incorporates the crucial role of plasma-molecule interactions (PMI) in the forward models of Balmer line emission. Through a mesh-based Bayesian inference paradigm, \divmidas{} utilises forward models of Langmuir probes, a divertor Thomson scattering system, and spectrally filtered camera images of Helium singlet and Hydrogen Balmer lines. Synthetic data sets, generated with realistic experimental errors from MAST-U Super-X SOLPS-ITER simulations, demonstrated \divmidas{}'s ability to accurately infer electron temperature and electron density fields under both attached and detached divertor conditions.

The analysis found that PMI must be accounted for in the Balmer line forward models in order to achieve agreement between the true (those used to produce the synthetic data set) and inferred plasma parameters. However, accounting for PMI through \qmol{} introduced additional complexity, increasing the uncertainty in the parameter combinations that could reproduce the data. This uncertainty predominantly stemmed from the potential misattribution of electron impact excitation emission to plasma-molecule interactions, resulting in over-estimation of \qmol{} and under-estimation of \ndens{}.

Furthermore, the study demonstrated the significant benefits of conducting inference on a mesh aligned to surfaces of constant poloidal magnetic flux. It was found that, even when combining multiple diagnostics, significantly diverse plasma parameter combinations could accurately replicate the data. However, the mesh allowed various spatially-dependent priors to be included in the inference. These helped to isolate physically plausible parameter combinations which, in turn, substantially improved the accuracy and precision of inferred plasma parameters. The approach achieved median absolute percentage errors within 6\% for electron temperature and 5\% for electron density across the divertor for detached and attached plasma conditions. This opens up the possibility of obtaining accurate two-dimensional \etemp{} and \edens{} inferences for MAST-U divertor experimental data.

\section{Acknowledgements}

Discussions with D. Moulton, N. Lonigro, T. van den Biggelaar, X. Pope, as well as the SOLPS-ITER simulations from A. Fil have kindly been acknowledged. This work has received support from EPSRC grants EP/T012250/1 and EP/N023846/1. This work has been carried out within the framework of the EUROfusion Consortium, partially funded by the European Union via the Euratom Research and Training Programme (Grant Agreement No 101052200 — EUROfusion). Views and opinions expressed are however those of the author(s) only and do not necessarily reflect those of the European Union or the European Commission. Neither the European Union nor the European Commission can be held responsible for them.

\newpage
\printbibliography

@article{hoffman2014no,
  title={The No-U-Turn sampler: adaptively setting path lengths in Hamiltonian Monte Carlo.},
  author={Hoffman, Matthew D and Gelman, Andrew and others},
  journal={J. Mach. Learn. Res.},
  volume={15},
  number={1},
  pages={1593--1623},
  year={2014}
}

@book{stangeby2000plasma,
  title={The plasma boundary of magnetic fusion devices},
  author={Stangeby, Peter C and others},
  volume={224},
  year={2000},
  publisher={Institute of Physics Pub. Philadelphia, Pennsylvania}
}

@article{Perek_2022,
doi = {10.1088/1741-4326/ac7813},
url = {https://dx.doi.org/10.1088/1741-4326/ac7813},
year = {2022},
month = {7},
publisher = {IOP Publishing},
volume = {62},
number = {9},
pages = {096012},
author = {A. Perek and M. Wensing and K. Verhaegh and B.L. Linehan and H. Reimerdes and C. Bowman and M. van Berkel and I.G.J. Classen and B.P. Duval and O. Février and J.T.W. Koenders and T. Ravensbergen and C. Theiler and M.R. de Baar and  the EUROfusion MST1 Team and  the TCV Team},
title = {A spectroscopic inference and SOLPS-ITER comparison of flux-resolved edge plasma parameters in detachment experiments on TCV},
journal = {Nuclear Fusion}
}

@misc{kingma2017adam,
      title={Adam: A Method for Stochastic Optimization}, 
      author={Diederik P. Kingma and Jimmy Ba},
      year={2017},
      eprint={1412.6980},
      archivePrefix={arXiv},
      primaryClass={cs.LG}
}

@misc{Liu1989,
      title={On the limited memory BFGS method for large scale optimization}, 
      author={Liu, Dong C. and Nocedal, Jorge},
      journal={Mathematical Programming},
      year={1989},
      volume={45},
      number={1},
      pages={503--528},
      doi = {10.1007/BF01589116},
      url = {https://doi.org/10.1007/BF01589116},
}

@article{neal2011mcmc,
  title={MCMC using Hamiltonian dynamics},
  author={Neal, Radford M and others},
  journal={Handbook of markov chain monte carlo},
  volume={2},
  number={11},
  pages={2},
  year={2011},
  publisher={Chapman and Hall/CRC}
}

@article{Fil_2022,
doi = {10.1088/1741-4326/ac81d8},
url = {https://dx.doi.org/10.1088/1741-4326/ac81d8},
year = {2022},
month = {8},
publisher = {IOP Publishing},
volume = {62},
number = {9},
pages = {096026},
author = {A. Fil and B. Lipschultz and D. Moulton and A. Thornton and B.D. Dudson and O. Myatra and K. Verhaegh and  the EUROfusion MST1 Team},
title = {Comparison between MAST-U conventional and Super-X configurations through SOLPS-ITER modelling},
journal = {Nuclear Fusion}
}

@article{Bowman_2020,
	doi = {10.1088/1361-6587/ab759b},
%	url = {https://doi.org/10.1088/1361-6587/ab759b},
	year = 2020,
	month = {2},
	publisher = {{IOP} Publishing},
	volume = {62},
	number = {4},
	pages = {045014},
	author = {C Bowman and J R Harrison and B Lipschultz and S Orchard and K J Gibson and M Carr and K Verhaegh and O Myatra},
	title = {Development and simulation of multi-diagnostic Bayesian analysis for 2D inference of divertor plasma characteristics},
	journal = {Plasma Physics and Controlled Fusion}
}

@Article{atoms4040026,
AUTHOR = {W\"underlich, Dirk and Fantz, Ursel},
TITLE = {Evaluation of State-Resolved Reaction Probabilities and Their Application in Population Models for He, H, and H2},
JOURNAL = {Atoms},
VOLUME = {4},
YEAR = {2016},
NUMBER = {4},
ARTICLE-NUMBER = {26},
URL = {https://www.mdpi.com/2218-2004/4/4/26},
ISSN = {2218-2004},
DOI = {10.3390/atoms4040026}
}

@online{adas,
  author = {ADAS Project and University of Strathclyde and the IAEA},
  title = {{OPEN}-{ADAS}: {Atomic Data and Analysis Structure}},
  year = 2018,
  url = {https://open.adas.ac.uk/},
  urldate = {2022-01-07}
}

@article{yacora,
title = {Yacora on the Web: Online collisional radiative models for plasmas containing H, H2 or He},
journal = {Journal of Quantitative Spectroscopy and Radiative Transfer},
volume = {240},
pages = {106695},
year = {2020},
issn = {0022-4073},
doi = {https://doi.org/10.1016/j.jqsrt.2019.106695},
%url = {https://www.sciencedirect.com/science/article/pii/S0022407319305163},
author = {D. Wünderlich and M. Giacomin and R. Ritz and U. Fantz},
}

@article{Lipschultz_2016,
	doi = {10.1088/0029-5515/56/5/056007},
	% url = {https://doi.org/10.1088/0029-5515/56/5/056007},
	year = 2016,
	month = {4},
	publisher = {{IOP} Publishing},
	volume = {56},
	number = {5},
	pages = {056007},
	author = {Bruce Lipschultz and Felix I. Parra and Ian H. Hutchinson},
	title = {Sensitivity of detachment extent to magnetic configuration and external parameters},
	journal = {Nuclear Fusion}
 }

@TechReport{amjuel,
author = {Reiter, Detlev},
year = {2000},
month = {01},
pages = {},
title = {The data file {AMJUEL}: Additional atomic and molecular data for {EIRENE}},
url = {http://www.eirene.de/html/amjuel.html},
institution   = {Forschungszentrum Jülich GmbH},
}

@article{WIESEN2015480,
title = {The new SOLPS-ITER code package},
journal = {Journal of Nuclear Materials},
volume = {463},
pages = {480-484},
year = {2015},
note = {PLASMA-SURFACE INTERACTIONS 21},
issn = {0022-3115},
doi = {https://doi.org/10.1016/j.jnucmat.2014.10.012},
url = {https://www.sciencedirect.com/science/article/pii/S0022311514006965},
author = {S. Wiesen and D. Reiter and V. Kotov and M. Baelmans and W. Dekeyser and A.S. Kukushkin and S.W. Lisgo and R.A. Pitts and V. Rozhansky and G. Saibene and I. Veselova and S. Voskoboynikov}
}

@article{Verhaegh_2023,
doi = {10.1088/1741-4326/acd394},
url = {https://dx.doi.org/10.1088/1741-4326/acd394},
year = {2023},
month = {05},
publisher = {IOP Publishing},
volume = {63},
number = {7},
pages = {076015},
author = {K. Verhaegh and A.C. Williams and D. Moulton and B. Lipschultz and B.P. Duval and O. Février and A. Fil and J. Harrison and N. Osborne and H. Reimerdes and C. Theiler and  the TCV Team and the EUROfusion MST1 Team},
title = {Investigating the impact of the molecular charge-exchange rate on detached SOLPS-ITER simulations},
journal = {Nuclear Fusion},
}

@article{Eich_2013,
doi = {10.1088/0029-5515/53/9/093031},
url = {https://dx.doi.org/10.1088/0029-5515/53/9/093031},
year = {2013},
month = {8},
publisher = {IOP Publishing and International Atomic Energy Agency},
volume = {53},
number = {9},
pages = {093031},
author = {T. Eich and A.W. Leonard and R.A. Pitts and W. Fundamenski and R.J. Goldston and T.K. Gray and A. Herrmann and A. Kirk and A. Kallenbach and O. Kardaun and A.S. Kukushkin and B. LaBombard and R. Maingi and M.A. Makowski and A. Scarabosio and B. Sieglin and J. Terry and A. Thornton and ASDEX Upgrade Team and JET EFDA Contributors},
title = {Scaling of the tokamak near the scrape-off layer H-mode power width and implications for ITER},
journal = {Nuclear Fusion},
}

@article{Hawke_2013,
doi = {10.1088/1748-0221/8/11/C11010},
url = {https://dx.doi.org/10.1088/1748-0221/8/11/C11010},
year = {2013},
month = {11},
publisher = {},
volume = {8},
number = {11},
pages = {C11010},
author = {J Hawke and  R Scannell and  J Harrison and  R Huxford and  P Bohm},
title = {Outline of optical design and viewing geometry for divertor Thomson scattering on MAST upgrade},
journal = {Journal of Instrumentation},
}

@article{Anderson1984,
title = {Simultaneous Algebraic Reconstruction Technique (SART): A superior implementation of the ART algorithm},
journal = {Ultrasonic Imaging},
volume = {6},
number = {1},
pages = {81-94},
year = {1984},
issn = {0161-7346},
doi = {https://doi.org/10.1016/0161-7346(84)90008-7},
url = {https://www.sciencedirect.com/science/article/pii/0161734684900087},
author = {A.H. Andersen and A.C. Kak},
}

@article{Bernert2023,
title = {The X-Point radiating regime at ASDEX Upgrade and TCV},
journal = {Nuclear Materials and Energy},
volume = {34},
pages = {101376},
year = {2023},
issn = {2352-1791},
doi = {https://doi.org/10.1016/j.nme.2023.101376},
url = {https://www.sciencedirect.com/science/article/pii/S2352179123000157},
author = {M. Bernert and S. Wiesen and O. Février and A. Kallenbach and J.T.W. Koenders and B. Sieglin and U. Stroth and T.O.S.J. Bosman and D. Brida and M. Cavedon and P. David and M.G. Dunne and S. Henderson and B. Kool and T. Lunt and R.M. McDermott and O. Pan and A. Perek and H. Reimerdes and U. Sheikh and C. Theiler and M. {van Berkel} and T. Wijkamp and M. Wischmeier},

}

@inproceedings{Lipschultz_1997,
  author={Lipschultz, B. and Goetz, J.A. and Hutchinson, I.H. and LaBombard, B. and McCracken, G.M. and Takase, Y. and Terry, J.L. and Bonou, P. and Golovato, S.N. and O'Shea, P. and Porkolab, M. and Boivin, R.L. and Bombarda, F. and Fiore, C.L. and Garnier, D.T. and Granetz, R.S. and Greenwald, M.J. and Horne, S.F. and Hubbard, A.E. and Irby, J.H. and Marmar, E.S. and May, M. and Mazurenko, A. and Reardon, J. and Rice, J.E. and Rost, C. and Schachter, J.M. and Snipes, J.A. and Stek, P.C. and Watterson, R.L. and Weaver, J. and Welch, B. and Wolfe, S.M.
  },
  booktitle={Proc. 16th Int. Conf. on Fusion Energy (Montreal, Canada)},
  volume={contributed papers, Part I},
  page={425},
  year={1996},
  organization={IAEA, Vienna}
}

@article{Moulton2023,
title = {Interpretive SOLPS-ITER simulations for MAST-U},
journal = {Nuclear Fusion, submitted},
author = {Moulton, D. and others},
year = {2024},
doi = {},
}

@article{Harrison2023,
  author        = {Harrison, J. and others},
  title         = {Overview of the MAST-Upgrade programme},
  booktitle     = {Nuclear Fusion, submitted},
  year            = {2024},
}

@article{Karhunen2023,
title = {Spectroscopic camera analysis of the roles of molecularly assisted reaction chains during detachment in JET L-mode plasmas},
journal = {Nuclear Materials and Energy},
volume = {34},
pages = {101314},
year = {2023},
issn = {2352-1791},
doi = {https://doi.org/10.1016/j.nme.2022.101314},
url = {https://www.sciencedirect.com/science/article/pii/S2352179122001958},
author = {J. Karhunen and A. Holm and S. Aleiferis and P. Carvalho and M. Groth and K.D. Lawson and B. Lomanowski and A.G. Meigs and A. Shaw and V. Solokha}
}

@article{Federici2023,
author = {Federici,Fabio  and Reinke,Matthew L.  and Lipschultz,Bruce  and Thornton,Andrew J.  and Harrison,James R.  and Lovell,Jack J.  and Bernert,Matthias },
title = {Design and implementation of a prototype infrared video bolometer (IRVB) in MAST Upgrade},
journal = {Review of Scientific Instruments},
volume = {94},
number = {3},
pages = {033502},
year = {2023},
doi = {10.1063/5.0128768}
}

@article{Pshenov2023,
title = {Divertor plasma opacity effects},
journal = {Nuclear Materials and Energy},
volume = {34},
pages = {101342},
year = {2023},
issn = {2352-1791},
doi = {https://doi.org/10.1016/j.nme.2022.101342},
url = {https://www.sciencedirect.com/science/article/pii/S235217912200223X},
author = {A.A. Pshenov and A.S. Kukushkin and A.V. Gorbunov and E.D. Marenkov}
}

@article{Linehan2023,
doi = {10.1088/1741-4326/acb5b0},
url = {https://dx.doi.org/10.1088/1741-4326/acb5b0},
year = {2023},
month = {2},
publisher = {IOP Publishing},
volume = {63},
number = {3},
pages = {036021},
author = {B.L. Linehan and A. Perek and B.P. Duval and F. Bagnato and P. Blanchard and C. Colandrea and H. De Oliveira and O. Février and E. Flom and S. Gorno and M. Goto and E. Marmar and L. Martinelli and A. Mathews and J. Muñoz-Burgos and D. Mykytchuk and N. Offeddu and D.S. Oliveira and H. Reimerdes and D. Reiter and O. Schmitz and J.L. Terry and C. Theiler and C.K. Tsui and B. Vincent and T. Wijkamp and C. Wüthrich and W. Zholobenko and the TCV Team},
title = {Validation of 2D ${{T}_\textrm{e}}$ and ${{n}_\textrm{e}}$ measurements made with Helium imaging spectroscopy in the volume of the TCV divertor},
journal = {Nuclear Fusion}
}

@article{Munoz2019,
    author = {Muñoz Burgos, J. M. and Griener, M. and Loreau, J. and Gorbunov, A. and Lunt, T. and Schmitz, O. and Wolfrum, E.},
    title = "{Evaluation of emission contributions from charge-exchange between the excited states of deuterium with He+ during diagnostic of thermal helium gas beam injection and laser-induced fluorescence}",
    journal = {Physics of Plasmas},
    volume = {26},
    number = {6},
    pages = {063301},
    year = {2019},
    month = {06},
    issn = {1070-664X},
    doi = {10.1063/1.5088363},
    url = {https://doi.org/10.1063/1.5088363},
    eprint = {https://pubs.aip.org/aip/pop/article-pdf/doi/10.1063/1.5088363/15973433/063301\_1\_online.pdf},
}

@article{Reimerdes2021,
doi = {10.1088/1741-4326/abd196},
url = {https://dx.doi.org/10.1088/1741-4326/abd196},
year = {2021},
month = {1},
publisher = {IOP Publishing},
volume = {61},
number = {2},
pages = {024002},
author = {H. Reimerdes and B.P. Duval and H. Elaian and A. Fasoli and O. Février and C. Theiler and F. Bagnato and M. Baquero-Ruiz and P. Blanchard and D. Brida and C. Colandrea and H. De Oliveira and D. Galassi and S. Gorno and S. Henderson and M. Komm and B. Linehan and L. Martinelli and R. Maurizio and J.-M. Moret and A. Perek and H. Raj and U. Sheikh and D. Testa and M. Toussaint and C.K. Tsui and M. Wensing and the TCV team and the EUROfusion MST1 team},
title = {Initial TCV operation with a baffled divertor},
journal = {Nuclear Fusion}
}

@PhdThesis{Biggelaar2022,
      title={Determination of a 2D electron density and electron temperature profile in the MAST-U divertor using spectral line ratios analysis}, 
      author={Thomas van den Biggelaar},
      year={2022},
  type          = {MSc Thesis},
  url           = {https://research.tue.nl/files/268853551/1003673_Biggelaar_T._van_den_MSc_thesis_Thesis_NF.pdf},
  school    = {Eindhoven University of Technology},
}

@article{Karhunen2020,
title = {Estimation of 2D distributions of electron density and temperature in the JET divertor from tomographic reconstructions of deuterium Balmer line emission},
journal = {Nuclear Materials and Energy},
volume = {25},
pages = {100831},
year = {2020},
issn = {2352-1791},
doi = {https://doi.org/10.1016/j.nme.2020.100831},
url = {https://www.sciencedirect.com/science/article/pii/S2352179120301022},
author = {J. Karhunen and B. Lomanowski and V. Solokha and S. Aleiferis and P. Carvalho and M. Groth and H. Kumpulainen and K.D. Lawson and A.G. Meigs and A. Shaw},
}

@article{Verhaegh_2022,
doi = {10.1088/1741-4326/aca10a},
url = {https://dx.doi.org/10.1088/1741-4326/aca10a},
year = {2022},
month = {12},
publisher = {IOP Publishing},
volume = {63},
number = {1},
pages = {016014},
author = {K. Verhaegh and B. Lipschultz and J.R. Harrison and N. Osborne and A.C. Williams and P. Ryan and J. Allcock and J.G. Clark and F. Federici and B. Kool and T. Wijkamp and A. Fil and D. Moulton and O. Myatra and A. Thornton and T.O.S.J. Bosman and C. Bowman and G. Cunningham and B.P. Duval and S. Henderson and R. Scannell and  the MAST Upgrade team},
title = {Spectroscopic investigations of detachment on the MAST Upgrade Super-X divertor},
journal = {Nuclear Fusion}
}

@misc{Verhaegh2023a,
      title={Investigating the impact of the molecular charge-exchange rate on detached SOLPS-ITER simulations}, 
      author={K. Verhaegh and Aelwyn C Williams and David Moulton and Bruce Lipschultz and Basil P. Duval and Olivier Fevrier and Alexandre Fil and Nick Osborne and Holger Reimerdes and Christian Theiler},
      year={2023},
      eprint={2301.11298},
      archivePrefix={arXiv},
      primaryClass={physics.plasm-ph},
}

@misc{Verhaegh2023c,
      title={Investigations of atomic \& molecular processes of NBI-heated discharges in the MAST Upgrade Super-X divertor}, 
      author={Kevin Verhaegh and others},
      year={2023},
      eprint={To be submitted},
}

@misc{Verhaegh2023d,
      title={Impact of Divertor Shape on Divertor Performance in strongly Baffled Divertors on MAST Upgrade}, 
      author={Kevin Verhaegh and others},
      year={2023},
      eprint={To be submitted},
}

@misc{Verhaegh2023b,
      title={The role of plasma-atom and molecule interactions on power \& particle balance during detachment on the MAST Upgrade Super-X divertor}, 
      author={Kevin Verhaegh and Bruce Lipschultz and James Harrison and Fabio Federici and David Moulton and Nicola Lonigro and Martin O'Mullane and Nick Osborne and Peter Ryan and Tijs Wijkamp and Bob Kool and Christian Theiler and Andrew Thornton},
      year={2023},
      eprint={2304.09109},
      archivePrefix={arXiv},
      primaryClass={physics.plasm-ph}
}

@article{Wijkamp2023,
doi = {10.1088/1741-4326/acc191},
url = {https://dx.doi.org/10.1088/1741-4326/acc191},
year = {2023},
month = {3},
publisher = {IOP Publishing},
volume = {63},
number = {5},
pages = {056003},
author = {T.A. Wijkamp and J.S. Allcock and X. Feng and B. Kool and B. Lipschultz and K. Verhaegh and B.P. Duval and J.R. Harrison and L. Kogan and N. Lonigro and A. Perek and P. Ryan and R.M. Sharples and I.G.J. Classen and R.J.E. Jaspers and the MAST Upgrade team},
title = {Characterisation of detachment in the MAST-U Super-X divertor using multi-wavelength imaging of 2D atomic and molecular emission processes},
journal = {Nuclear Fusion}
}

@article{Kuang2020, 
    title={Divertor heat flux challenge and mitigation in SPARC}, 
    volume={86}, 
    DOI={10.1017/S0022377820001117}, 
    number={5}, 
    journal={Journal of Plasma Physics}, publisher={Cambridge University Press}, author={Kuang, A. Q. and Ballinger, S. and Brunner, D. and Canik, J. and Creely, A. J. and Gray, T. and Greenwald, M. and Hughes, J. W. and Irby, J. and LaBombard, B. and et al.},
    year={2020}, 
    pages={865860505}
}

@article{Feng2021,
author = {Feng,X.  and Calcines,A.  and Sharples,R. M.  and Lipschultz,B.  and Perek,A.  and Vijvers,W. A. J.  and Harrison,J. R.  and Allcock,J. S.  and Andrebe,Y.  and Duval,B. P.  and Mumgaard,R. T. },
title = {Development of an 11-channel multi wavelength imaging diagnostic for divertor plasmas in MAST Upgrade},
journal = {Review of Scientific Instruments},
volume = {92},
number = {6},
pages = {063510},
year = {2021},
doi = {10.1063/5.0043533},

URL = { 
        https://doi.org/10.1063/5.0043533
    
},
eprint = { 
        https://doi.org/10.1063/5.0043533
    
}

}

@article{Verhaegh2021b,
	doi = {10.1088/1741-4326/ac1dc5},
	url = {https://doi.org/10.1088/1741-4326/ac1dc5},
	year = 2021,
	month = {9},
	publisher = {{IOP} Publishing},
	volume = {61},
	number = {10},
	pages = {106014},
	author = {K. Verhaegh and B. Lipschultz and J.R. Harrison and B.P. Duval and A. Fil and M. Wensing and C. Bowman and D.S. Gahle and A. Kukushkin and D. Moulton and A. Perek and A. Pshenov and F. Federici and O. F{\'{e}}vrier and O. Myatra and A. Smolders and C. Theiler and  the TCV Team and  the EUROfusion MST1 Team},
	title = {The role of plasma-molecule interactions on power and particle balance during detachment on the {TCV} tokamak},
	journal = {Nuclear Fusion}
}

@article{Perek2021,
title = {Measurement of the 2D emission profiles of hydrogen and impurity ions in the TCV divertor},
journal = {Nuclear Materials and Energy},
volume = {26},
pages = {100858},
year = {2021},
issn = {2352-1791},
doi = {10.1016/j.nme.2020.100858},
author = {A. Perek and B.L. Linehan and M. Wensing and K. Verhaegh and I.G.J. Classen and B.P. Duval and O. Février and H. Reimerdes and C. Theiler and T.A. Wijkamp and M.R. {de Baar}}
}

@article{Pitts2019,
title = {Physics basis for the first ITER tungsten divertor},
journal = {Nuclear Materials and Energy},
volume = {20},
pages = {100696},
year = {2019},
issn = {2352-1791},
doi = {https://doi.org/10.1016/j.nme.2019.100696},
url = {https://www.sciencedirect.com/science/article/pii/S2352179119300237},
author = {R.A. Pitts and X. Bonnin and F. Escourbiac and H. Frerichs and J.P. Gunn and T. Hirai and A.S. Kukushkin and E. Kaveeva and M.A. Miller and D. Moulton and V. Rozhansky and I. Senichenkov and E. Sytova and O. Schmitz and P.C. Stangeby and G. {De Temmerman} and I. Veselova and S. Wiesen}
}

@TechReport{Reiter2018,
title = {Isotope effects in molecule assisted recombination and dissociation in divertor plasmas},
series = {Berichte des Forschungszentrums Jülich ;},
author = {Janev, R. K. and Reiter, D.},
address = {Jülich},
publisher = {Forschungszentrum, Zentralbibliothek},
year = {2018},
pages = {1 Online-Ressource (37 Seiten)},
note = {englisch},
type = {J\"{u}lich report - JUEL 4411},
url = {https://juser.fz-juelich.de/record/850290/files/J\%C3\%BCl\_4411\_Reiter.pdf?version=1},
institution   = {Forschungszentrum Jülich GmbH}
}

@article{Wenninger2014,
	doi = {10.1088/0029-5515/54/11/114003},
	year = 2014,
	month = {11},
	publisher = {{IOP} Publishing},
	volume = {54},
	number = {11},
	pages = {114003},
	author = {R.P. Wenninger and M. Bernert and T. Eich and E. Fable and G. Federici and A. Kallenbach and A. Loarte and C. Lowry and D. McDonald and R. Neu and others},
	title = {{DEMO} divertor limitations during and in between {ELMs}},
	journal = {Nuclear Fusion}
}

@article{Verhaegh2021,
	doi = {10.1088/1361-6587/abd4c0},
	year = 2021,
	month = {1},
	publisher = {{IOP} Publishing},
	volume = {63},
	number = {3},
	pages = {035018},
	author = {K. Verhaegh and B Lipschultz and C Bowman and B P Duval and U Fantz and A Fil and J R Harrison and D Moulton and O Myatra and D Wünderlich and F Federici and D S Gahle and A Perek and M Wensing and   and},
	title = {A novel hydrogenic spectroscopic technique for inferring the role of plasma{\textendash}molecule interaction on power and particle balance during detached conditions},
	journal = {Plasma Physics and Controlled Fusion}
}

@article{Verhaegh2021a,
title = {A study on the influence of plasma-molecule interactions on particle balance during detachment},
author = {K Verhaegh and B Lipschultz and J R Harrison and B P Duval and C Bowman and A Fil and D S Gahle and D Moulton and O Myatra and A Perek and C Theiler and M Wensing},
journal = {Nuclear Materials and Energy},
year = {2021},
doi = {10.1016/j.nme.2021.100922},
volume = {26},
pages = {100922}
}

@article{Lawson_2024,
doi = {10.1088/1361-6587/ad75b9},
url = {https://dx.doi.org/10.1088/1361-6587/ad75b9},
year = {2024},
month = {9},
publisher = {IOP Publishing},
volume = {66},
number = {11},
pages = {115001},
author = {K D Lawson and I H Coffey and M Groth and A G Meigs and S Menmuir and B Thomas and JET Contributors},
title = {He II line intensity measurements in the JET tokamak},
journal = {Plasma Physics and Controlled Fusion}
}

@article{Lomanowski2020,
  title={Interpretation of Lyman opacity measurements in {JET with the ITER-like wall} using a particle balance approach},
  author={Lomanowski, B. and Groth, Mathias and Coffey, Ivor H and Karhunen, Juuso and Maggi, Costanza F and Meigs, Andy and Menmuir, Sheena and O'Mullane, Martin},
  journal={Plasma Physics and Controlled Fusion},
  year={2020},
  publisher={IOP Publishing},
  doi = {10.1088/1361-6587/ab7432},
  volume={62},
  number={065006}
}

@Article{Feng2017,
  author        = {Feng, Y. and Frerichs, H. and Kobayashi, M. and Reiter, D.},
  title         = {Monte-Carlo fluid approaches to detached plasmas in non-axisymmetric divertor configurations},
  journal       = {Plasma Physics and Controlled Fusion},
  year          = {2017},
  volume        = {59},
  number        = {3},
  pages         = {034006},
  issn          = {0741-3335},
  __markedentry = {[kevin:6]},
  type          = {Journal Article},
  url           = {http://stacks.iop.org/0741-3335/59/i=3/a=034006},
}

@Article{Havlickova2015,
  author        = {Havlickova, E. and Harrison, J. and Lipschultz, B. and Fishpool, G. and Kirk, A. and Thornton, A. and Wischmeier, M. and Elmore, S. and Allan, S.},
  title         = {SOLPS analysis of the MAST-U divertor with the effect of heating power and pumping on the access to detachment in the Super-x configuration},
  journal       = {Plasma Physics and Controlled Fusion},
  year          = {2015},
  volume        = {57},
  number        = {11},
  pages         = {115001},
  issn          = {0741-3335},
  __markedentry = {[kevin:6]},
  doi           = {Artn 115001
10.1088/0741-3335/57/11/115001},
  type          = {Journal Article},
  url           = {<Go to ISI>://WOS:000374538100002
http://iopscience.iop.org/article/10.1088/0741-3335/57/11/115001/pdf},
}

@Article{Henderson2018,
  author        = {Henderson, S. S. and Bernert, M. and Brezinsek, S. and Carr, M. and Cavedon, M. and Dux, R. and Lipschultz, B. and O’Mullane, M. G. and Reimold, F. and Reinke, M. L. and The, Asdex Upgrade Team and The, M. S. T. Team},
  title         = {Determination of volumetric plasma parameters from spectroscopic N II and N III line ratio measurements in the ASDEX Upgrade divertor},
  journal       = {Nuclear Fusion},
  year          = {2018},
  volume        = {58},
  number        = {1},
  pages         = {016047},
  issn          = {0029-5515},
  __markedentry = {[kevin:6]},
  type          = {Journal Article},
  url           = {http://stacks.iop.org/0029-5515/58/i=1/a=016047},
}

@Article{Kukushkin2017,
  author        = {Kukushkin, A. S. and Krasheninnikov, S. I. and Pshenov, A. A. and Reiter, D.},
  title         = {Role of molecular effects in divertor plasma recombination},
  journal       = {Nuclear Materials and Energy},
  year          = {2017},
  volume        = {12},
  pages         = {984-988},
  issn          = {2352-1791},
  __markedentry = {[kevin:6]},
  doi           = {10.1016/j.nme.2016.12.030},
  type          = {Journal Article},
}

@article{Allen1997,
title = {First measurements of electron temperature and density with divertor Thomson scattering in radiative divertor discharges on DIII-D},
journal = {Journal of Nuclear Materials},
volume = {241-243},
pages = {595-601},
year = {1997},
issn = {0022-3115},
doi = {https://doi.org/10.1016/S0022-3115(97)80106-7},
url = {https://www.sciencedirect.com/science/article/pii/S0022311597801067},
author = {S.L. Allen and D.N. Hill and T.N. Carlstrom and D.G. Nilson and R. Stockdale and C.L. Hsieh and T.W. Petrie and A.W. Leonard and D. Ryutov and G.D. Porter and R. Maingi and M.R. Wade and R. Cohen and W. Nevins and M.E. Fenstermacher and R.D. Wood and C.J. Lasnier and W.P. West and M.D. Brown}
}

@Article{Lipschultz2007a,
  author        = {Lipschultz, B. and LaBombard, B. and Terry, J. L. and Boswell, C. and Hutchinson, I. H.},
  title         = {Divertor physics research on {Alcator C-Mod}},
  journal       = {Fusion Science and Technology},
  year          = {2007},
  volume        = {51},
  number        = {3},
  pages         = {369-389},
  issn          = {1536-1055},
  __markedentry = {[kevin:6]},
  type          = {Journal Article},
}

@article{Perek2019submitted,
author = {Perek,A.  and Vijvers,W. A. J.  and Andrebe,Y.  and Classen,I. G. J.  and Duval,B. P.  and Galperti,C.  and Harrison,J. R.  and Linehan,B. L.  and Ravensbergen,T.  and Verhaegh,K.  and de Baar,M. R. },
title = {MANTIS: A real-time quantitative multispectral imaging system for fusion plasmas},
journal = {Review of Scientific Instruments},
volume = {90},
number = {12},
pages = {123514},
year = {2019},
doi = {10.1063/1.5115569},
}

@Article{Reimold2015,
  author        = {Reimold, F. and Wischmeier, M. and Bernert, M. and Potzel, S. and Kallenbach, A. and Muller, H. W. and Sieglin, B. and Stroth, U. and Team, ASDEX Upgrade},
  title         = {Divertor studies in nitrogen induced completely detached H-modes in full tungsten ASDEX Upgrade},
  journal       = {Nuclear Fusion},
  year          = {2015},
  volume        = {55},
  number        = {3},
  pages         = {033004},
  issn          = {0029-5515},
  __markedentry = {[kevin:6]},
  doi           = {Artn 033004
10.1088/0029-5515/55/3/033004},
  type          = {Journal Article},
}

@Article{Stangeby1993,
  author        = {Stangeby, P. C.},
  title         = {Can detached divertor plasmas be explained as self-sustained gas targets?},
  journal       = {Nuclear fusion},
  year          = {1993},
  volume        = {33},
  number        = {11},
  pages         = {1695},
  __markedentry = {[kevin:6]},
  type          = {Journal Article},
}

@Article{Theiler2017,
  author        = {Theiler, C. and Lipschultz, B. and Harrison, J. and Labit, B. and Reimerdes, H. and Tsui, C. and Vijvers, W. A. J. and Boedo, J. A. and Duval, B. P. and Elmore, S. and Innocente, P. and Kruezi, U. and Lunt, T. and Maurizio, R. and Nespoli, F. and Sheikh, U. and Thornton, A. J. and van Limpt, S. H. M. and Verhaegh, K. and Vianello, N. and Team, TCV and Team, EUROfusion MST1},
  title         = {Results from recent detachment experiments in alternative divertor configurations on TCV},
  journal       = {Nuclear Fusion},
  year          = {2017},
  volume        = {57},
  number        = {7},
  pages         = {072008},
  issn          = {0029-5515},
  __markedentry = {[kevin:6]},
  doi           = {ARTN 072008
10.1088/1741-4326/aa5fb7},
  type          = {Journal Article},
  url           = {http://iopscience.iop.org/article/10.1088/1741-4326/aa5fb7/pdf},
}

@Article{Verhaegh2019,
  author        = {Verhaegh, K. and Lipschultz, Bruce and Duval, Basil and Février, Olivier and Fil, Alexandre and Theiler, Christian and Wensing, Mirko and Bowman, Christopher and Gahle, Daljeet and Harrison, James and others},
  title         = {An improved understanding of the roles of atomic processes and power balance in divertor target ion current loss during detachment},
  journal       = {Nuclear Fusion},
  year          = {2019},
  volume        = {59},
  number        = {126038},
  doi           = {10.1088/1741-4326/ab4251},
  type          = {Journal Article},
}

@Article{Wuenderlich2016,
  author        = {Wünderlich, Dirk and Fantz, Ursel},
  title         = {Evaluation of State-Resolved Reaction Probabilities and Their Application in Population Models for {He, H, and H2}},
  journal       = {Atoms},
  year          = {2016},
  volume        = {4},
  number        = {4},
  issn          = {2218-2004},
  __markedentry = {[kevin:6]},
  doi           = {10.3390/atoms4040026},
  type          = {Journal Article}
}

@article{Hudoba2023,
  title={Magnetic equilibrium optimisation and divertor integration in spherical tokamak reactors},
  author={Hudoba, A. and Cunningham, G and Bakes, S and STEP Team and others},
  journal={Fusion Engineering and Design},
  volume={191},
  pages={113704},
  year={2023},
  publisher={Elsevier}
}

@article{Osawa2023,
doi = {10.1088/1741-4326/acd863},
url = {https://dx.doi.org/10.1088/1741-4326/acd863},
year = {2023},
month = {6},
publisher = {IOP Publishing},
volume = {63},
number = {7},
pages = {076032},
author = {R.T. Osawa and D. Moulton and S.L. Newton and S.S. Henderson and B. Lipschultz and A. Hudoba},
title = {SOLPS-ITER analysis of a proposed STEP double null geometry: impact of the degree of disconnection on power-sharing},
journal = {Nuclear Fusion},
}

\newpage
\begin{appendices}
\section{Emission Models}
\subsection{Molecular Emission Model}
\label{sec:Appendix: Molecular emission model}
Most Balmer line emission arising from molecular break-up in the divertor occurs when molecular ions (mostly $D_2^+$, but also $D_2^- \rightarrow D^- + D$ and $D_3^+$) interact with the plasma \cite{atoms4040026}. Modeling the Balmer line emission from plasma-molecule interactions requires the photon-emissivity coefficients \footnote{Akin to those provided by ADAS for atomic processes as used in equation\ref{eqn:emissivity}.} for each process, $\pec{}_{D_2}$, $\pec{}^{n\to 2}_{D_2^+}$, $\pec{}^{n\to 2}_{D^-}$, $\pec{}^{n\to 2}_{D_3^+}$,  as well as each of the respective molecular densities.

Since each field required by the forward models introduces $V$ additional parameters to our inference, the inclusion of these density fields would lead to a costly increase in our parameter space. To reduce the number of free parameters required to include molecular effects, an effective molecular photo-emissivity coefficient is derived, 
\begin{equation}
    \label{eqn:PEC_D2eff}
    \eqalign{\pec{}_{\dtwoeff{}}^{n\rightarrow2} (T_e, n_e) = \pec{}_{D_2}^{n\rightarrow2} + \frac{n_{D_2^+}}{n_{D_2}} \pec{}_{D_2^+}^{n\rightarrow2} + \cr\ \frac{n_{D^-}}{n_{D_2}} \pec{}_{D^-}^{n\rightarrow2}  + \frac{n_{D_3^+}}{n_{D_2}} \pec{}_{D_3^+}^{n\rightarrow2}.} 
\end{equation}
This encapsulates the molecular contribution to the Balmer-line emissivity (with each $\pec{}$ term depending on \edens{} and \etemp{} alone) whilst avoiding the need to include additional parameters for each molecular process in equation~(\ref{eqn:emissivity}). 

Under the assumption of no transport, the ratios of the various molecular ion densities to the molecular density \footnote{assuming the electron density and the hydrogen ion density are roughly equal} in equation~(\ref{eqn:PEC_D2eff}) can be estimated by the ratio between the creation and destruction rates of each molecular species \cite{Verhaegh2021a,Verhaegh2021b}. These creation and destruction rates were adopted from the default hydrogenic rates used by Eirene \cite{amjuel} as a function of \etemp{} and \edens{} \cite{amjuel}. These rates are for hydrogen, but we aim to analyse data of a deuterium plasma. As such, translation factors for hydrogen to deuterium, adopted from \cite{Reiter2018}, were applied ($0.95$ and $0.7$ for $\frac{n_{D_2^+}}{n_{D_2}}$ and $\frac{n_{D^-}}{n_{D_2}}$ respectively), rather than using an isotope mass-based rescaling, which exacerbates errors in the molecular charge-exchange rates at low temperatures \cite{Verhaegh2023a}. The respective $\pec{}$ values for plasma-molecular interactions were obtained from the population coefficients retrieved from the YACORA-on-the-web \cite{yacora,Wuenderlich2016} collisional-radiative model. In detachment-relevant conditions (0.6-2 eV), $D_2^+$ is the dominant driver of hydrogen emission associated with plasma-molecular break-up: $\pec{}_{\dtwoeff{}}^{n\rightarrow2} (T_e, n_e) \approx \frac{n_{D_2^+}}{n_{D_2}} \pec{}_{D_2^+}^{n\rightarrow2}$.

The uncertainties in the relative ratios of molecular ion densities to $D_2$ in equation~(\ref{eqn:PEC_D2eff}) are large \cite{Verhaegh_2023}. There are strong concerns about the various creation rates for molecular ions that Eirene employs, particularly $D_2 + D^+ \rightarrow D_2^+ + D$ which is the dominant driver for $D_2^+$ in detached conditions \cite{Verhaegh2021b,Verhaegh_2022,Verhaegh2023a,Verhaegh2023b,Verhaegh2023c,Kukushkin2017}. Molecular ion creation is driven by vibrationally excited molecules, which depend not only on local \edens{} and \etemp{}, but also on molecular transport and plasma-wall interactions. Additionally, there are strong concerns about the vibrational distribution model employed by Eirene \cite{Verhaegh2021b,Verhaegh2023a,Verhaegh2023c}. These concerns lead to uncertainties of multiple orders of magnitude in the amplitude of $\pec{}_{\dtwoeff{}}^{n\rightarrow2}$.

If only one emission process is dominant, \emph{the ratio between the emission coefficient of two Balmer lines is not affected by these uncertainties}
. Therefore, rather than multiplying $\pec{}_{\dtwoeff{}}^{n\to 2}$ by \edens{} and the neutral molecular density $n_{D_2}$, we express our emission model in such a way that only the ratios of $\pec{}_{\dtwoeff{}}^{n\rightarrow2}$ between different Balmer lines are used. This re-parameterisation utilises \qmol{} which is defined to be the ratio of $D_\alpha$'s molecular emissivity contributions, $\varepsilon^{3\to 2}_{\mathrm{mol}}$, to atomic emissivity contributions, $\varepsilon^{3\to 2}_{\mathrm{atm}}$,
\begin{equation}
    Q_{D_\alpha}^{\mathrm{mol}} \equiv \frac{\varepsilon^{3\to 2}_{\mathrm{mol}}}{\varepsilon^{3\to 2}_{\mathrm{atm}}} = \frac{\varepsilon^{3\to 2}_{\dtwoeff{}}}{\varepsilon^{3\to 2}_{\mathrm{rec}} + \varepsilon^{3\to 2}_{\mathrm{exc}}}.
\end{equation}
The emissivity of Balmer lines $n\to2$ were thus given by,
\begin{equation}
    \eqalign{
    \varepsilon_{n\to 2} = \underbrace{n_e^2 \pec{}^{n\to 2}_{rec} + n_Hn_e \pec{}^{n\to 2}_{\mathrm{exc}}}_{\mathrm{Atomic\; contribution,}\; \varepsilon^{n\to2}_{\mathrm{atm}}} +\cr\ \underbrace{Q_{D_\alpha}^{\mathrm{mol}}\left(n_e^2 \pec{}^{3\to 2}_{\mathrm{rec}} + n_Hn_e \pec{}^{3\to 2}_{\mathrm{exc}}\right)\frac{\pec{}_{\dtwoeff{}}^{n\to 2}}{\pec{}_{\dtwoeff{}}^{3\to 2}}}_{\mathrm{Molecular\; contribution,}\; \varepsilon^{n\to2}_{\mathrm{mol}}} .}
\end{equation}
Consequently, no molecular density was explicitly required. Furthermore, the effective molecular PEC was only present as a ratio between lines $n\to 2$ and $3\to 2$ minimising the impact of the aforementioned approximations. This permitted molecular processes to be included by a single field parameter with minimally impactful approximations.

\subsection{Helium Emission Models}
\label{sec:Appendix: Helium details}
MAST-U's MWI and TCV's MANTIS cameras also routinely capture Helium-I emission lines which has been shown \cite{Linehan2023} to provide information on \edens{} and \etemp{}. The nature of integrated data analysis is such that a single inaccurate forward model can offset the overall inference. As such we explicitly include parameters for \hendens{} and \hepdens{} at each mesh vertex (at a cost of 2V additional free parameters). Doing so avoids assumptions of ionisation balance models and permits inclusion of both electron-impact excitation (of $\mathrm{He}^0$) and electron-ion recombination (of $\mathrm{He}^+$) contributions to emission. He I PECs were provided by ADAS \cite{adas} and implemented as with equation~(\ref{eqn:emissivity}). 

In conditions with significant hydrogen atomic densities, charge-exchange between helium and hydrogen could occur. This can change both the ionisation fractions of $\mathrm{He}$ (which is included in our model through the \hendens{} and \hepdens{} parameters) and lead to the emission of He I \cite{Munoz2019}. Although this is neglected in our emission model, since population coefficients for this interaction are not freely available, having neutral hydrogen density, \ndens{}, as a parameter in our model means that charge exchange can trivially be included once population coefficients for this interaction become available. 

The 668, 728 and 502 nm singlet He I lines were chosen for the helium forward models. This followed from the work by \cite{Biggelaar2022}, which used these transitions due to their insensitivity to transport of metastable states, which can impact singlet to triplet He I line ratios. Detailed studies on TCV using He I emission \cite{Linehan2023} have suggested that the magnetic field as well as molecular effects could impact He I emission. However, based on those results, and the relatively low magnetic field at MAST-U (0.7 T on-axis), we would not expect that this would impact the chosen He I singlet transitions significantly.

\section{Complications to the inference}
\subsection{Challenges associated with Q-mol inferences and possible improvements}

In Section~\ref{sec:fullIDA-result} we have observed that the \ndens{} inferences are associated with large uncertainties. Since \ndens{} and \qmol{} depend mainly on neutral atom and molecular densities, they are not necessarily field-aligned. Therefore, only very weak/no spatially-dependent priors could be applied to the \ndens{} / \qmol{} parameters. 

We have observed that there is a negative correlation in the samples of \ndens{} and \qmol{} (Figure~\ref{fig:MEX Scatter} c, d). As shown in equation~(\ref{eqn:Balmer_emissivity}), both \ndens{} and \qmol{} can be freely altered to match the required (i.e., synthetically measured) emissivity. This suggests that the emission problem is ill-posed for these parameters, which is exacerbated by the lack of strong priors for \qmol{}. 

The ability of the analysis to detect PMI related contributions to the Balmer line emission is further investigated in Figure~\ref{fig:EmissivityFractions}. Here, the true $D_\gamma$ emissivity profile, split by the different emission processes, in the \state{1} is compared with the inferred one. The general structure in the MWI synthetic diagnostic observation region is well represented. Outside the MWI synthetic diagnostic observation region, however, both the total emission as well as its PMI contribution is over-predicted due to a lack of constraint on the total emission arising from that region. Despite this over-prediction of PMI emission, \etemp{}, \edens{} and \ndens{} are still reasonably well inferred, which is attributed to the spatially-dependent priors.

Although the spatially-dependent priors are crucial to the IDA (see section~\ref{sec:Importance of Spatially-Dependent Priors}), they can result in slight (within uncertainty) discrepancies between the forward-modelled image and the data of the synthetic diagnostic. For example, \ndens{} (\qmol{}) are overestimated near R=1.2 m, Z=-1.8 m (Figure~\ref{fig:EmissivityFractions}, \ref{fig:Results} a.iii.). In this region, the flexibility of \qmol{}, combined with the ill-posed nature of the emission model and the favouritism towards smooth solutions create a `perfect storm'; the posterior probability of this incorrect solution, where PMI is overestimated, is higher than the `known' solution.

The inferences of \qmol{} could be improved through: introducing priors on the \qmol{} profile; or the inclusion of additional diagnostics that provide information on \ndens{} and/or \qmol{}. This information can be provided indirectly through techniques that enable separating EIE and PMI hydrogen emission. One possibility is to utilise the $D_2$ Fulcher emission that has been successfully applied to line-integrated spectroscopy \cite{Verhaegh2023c}. An additional Balmer-line camera in the upstream region, such as the X-Point imaging camera recently installed on MAST-U, could be incorporated into the IDA to improve the inferences in the region where the MWI coverage is limited.

\subsection{Numerical complexities of characterising the Posterior Distribution}
\label{sec:Appendix: compl_posterior}
\begin{figure*}
    \centering
    \includegraphics[width=\linewidth]{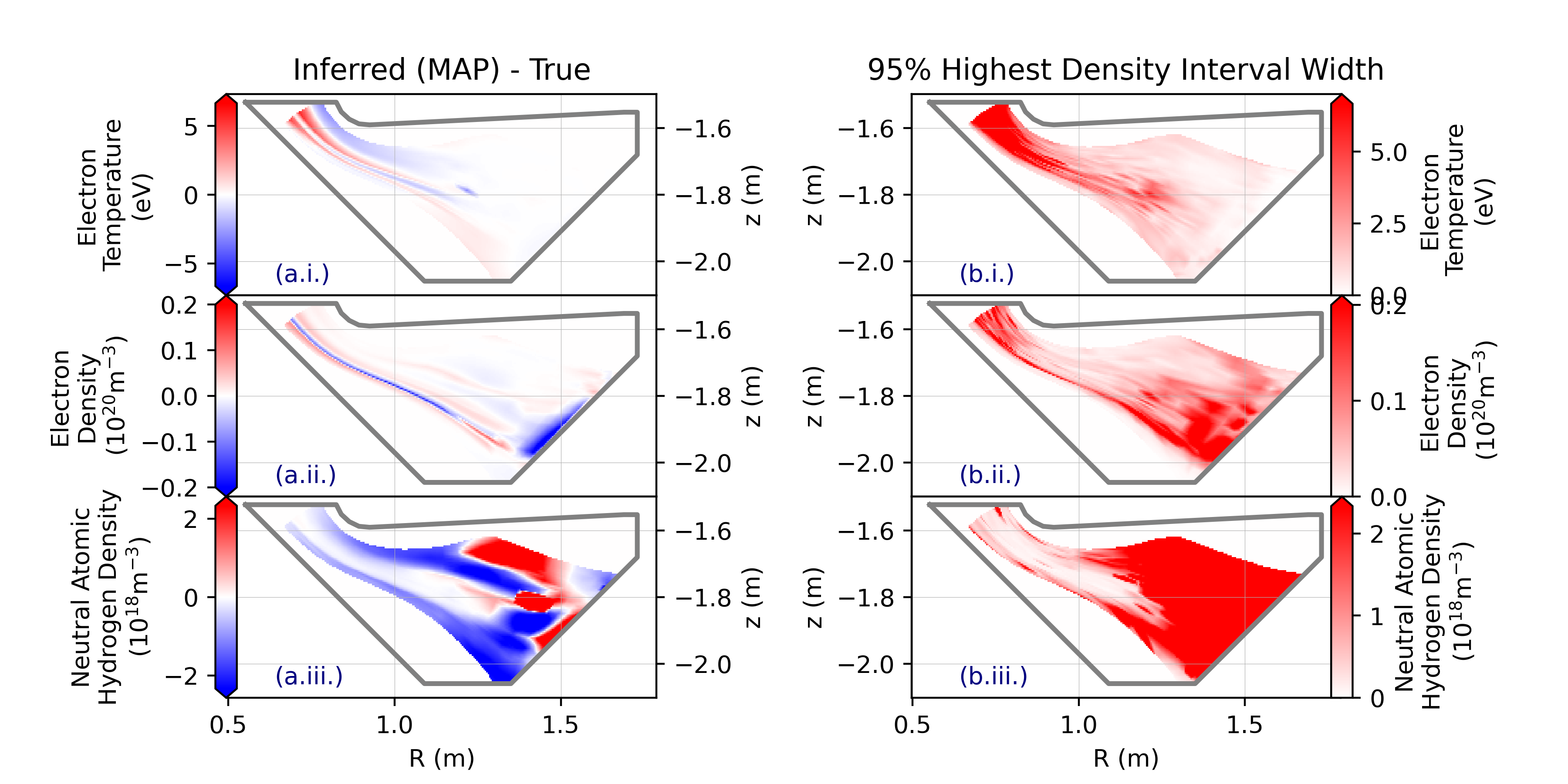}
    \caption{Visualisation of error (a) and uncertainty (b) for the detached case using posterior (i).}
    \label{fig:uncertainty_quantification}
\end{figure*}

The nature of our posterior distribution posed three key challenges: the number of parameters, the correlations between parameters and multi-modality. These challenges especially hamper adequate sampling of the posterior distribution. Hamiltonian Monte Carlo (HMC) \cite{neal2011mcmc}, with a mass matrix constructed from an approximation of the Hessian in the MAP vicinity countered these challenges. Figure \ref{fig:uncertainty_quantification} displays the error in our inferred parameters (a) and the inferred uncertainty (b). We find the regions of larger error do indeed correspond to increased inferred uncertainty. This suggests that, despite the complex posterior distribution, HMC was able to adequately characterise it.

The HMC algorithm required significant computational resources. This was due to the small time-step necessary to avoid divergence in the HMC algorithm and the accompanying large number of leap-frog steps necessary to avoid inefficient random-walk like behaviour. This is observed in this work in figure~\ref{fig:MEX Scatter} where only 63\% of \ndens{} and \hepdens{} parameters contain the true value within their 95\% HDI suggesting incomplete sampling. An improved method for finding the mass matrix used in the HMC algorithm is required for more efficient sampling.

The finding of the MAP, using the genetic algorithm outlined in the appendix of~\cite{Bowman_2020}, was negatively impacted by the use of the Cauchy prior probability function. The heavier tails, in comparison to the Gaussian distribution, made for less-pronounced peaks in the probability distribution. This increased the number of local-maxima far from the global maximum resulting in a longer time for the genetic algorithm to reach convergence. This was somewhat countered by the hybrid use of the momentum-based, first-order adam optimiser \cite{kingma2017adam} and the second-order L-BFGS optimiser \cite{Liu1989} which carry different benefits in different regions of the parameter space.

The significantly more computational time demanded by a mesh-based IDA over a cell-based approach (with no more than six free-parameters, see Section~\ref{sec:ida setup}) prevents its use for inter-shot analysis. The IDA, as illustrated here, is envisaged to obtain more accurate estimates for a reduced set of cases that require analysing in detail. Cell-based approaches can be used as a more general approach that can be used for standard, and potentially inter-shot, analysis.

\subsection{Improving the analysis through spatially-dependent priors and its limitations}
As outlined in Section~\ref{sec:Importance of Spatially-Dependent Priors}, the spatially dependent priors introduced in this work significantly improved the inference both in terms of precision (reducing its inferred uncertainty) and accuracy (an improved match between the `known' result and the MAP estimate). The most effective of these priors were those on the smoothness of the parameter fields. 

Figure 6 of \cite{Bowman_2020} outlined how varying the `strength' (the reciprocal of the $\sigma$ parameter for a Gaussian distribution) of the smoothing prior probability distribution can significantly alter the inference. Investigations found that each plasma state (i.e. the \state{1} and the \state{2}) had different optimal smoothing strengths that cannot be known \textit{a priori}. 

The benefit of the empirical approach to selecting smoothing strengths, as outlined in Section~\ref{sec:Appendix: Prior Distributions}, came from averaging over a range of different divertor states. This demanded a heavy-tailed Cauchy distribution which did not penalise non-smooth, sharp ridges (as was found for some plasma states) as harshly as the light-tailed Gaussian distribution. Consequently, the inference became less-sensitive to the choice of smoothing strength, meaning that a single smoothing strength value (per prior) could be selected and applied to inferences of all relevant plasma states of MAST-U.

The choice of a Cauchy distribution to handle the smoothing priors came with limitations. Diagnostic noise could suggest a sharp feature in the plasma state which a Cauchy smoothing prior could incorrectly permit. This can be seen by the un-smooth drop in electron temperature in the region of (1.2 -1.7)m of figures~\ref{fig:Results} and~\ref{fig:EmissivityFractions}. 
Consequently, when moving to experimental data, a more sophisticated approach to handling the crucial smoothing priors may be required.

\section{Priors}
\label{sec:Appendix: Priors}
\subsection{Prior distributions}
\label{sec:Appendix: Prior Distributions}

Each prior requires: the calculation, from $\theta$, of the physical quantity that the prior contribution is describing (the `argument'), $f_j(\theta)$; and a probability function describing the expected distribution of these arguments. Spatially-dependent priors on the smoothness of the parameter fields are particularly useful in our work. However, whilst a smooth field is more probable than a noisy field, perfectly smooth fields are not expected. This highlights the importance of the choice of the (log) prior probability function; the probability to attribute to different degrees of smoothness (quantified as the prior argument) is somewhat arbitrary but significantly impacts the inference.

An empirical approach was used to determine the prior probability functions. For each prior, its argument at each mesh vertex was evaluated across many SOLPS-ITER simulations (excluding those used to generate the synthetic data in this work). This resulted in a distribution of expected prior arguments for each prior. By fitting these distributed prior arguments with different differentiable probability distributions, we found that a heavy-tailed function was optimal. This allowed for a preference for compliance with the prior rationale without significantly penalising outliers. For example, the majority of vertices in the mesh would be expected to have a smooth variation in \edens{} parallel to flux surfaces but, by having a heavy-tailed distribution, sharp variations are permitted should the data suggest so. 

 For each prior, it was found that a Cauchy distribution best characterised the prior arguments (with each mesh vertex, $k$, independently treated). Consequently, each prior had a log-probability of
\begin{equation}
    \mathcal{L}_j(\thetavec) = V\log(\gamma_j) - \sum_{k=1}^{V}\log\left(\left(f_j\left(\theta_k,...\right)\right)^2 + \gamma_j^2\right).
\end{equation}
$\gamma_j$, the scale parameter, was empirically found for each individual prior and details of these are outlined in Appendix~\ref{sec:Appendix: Prior details}.

\subsection{Prior details}
\label{sec:Appendix: Prior details}
\noindent
For each prior, its argument was evaluated at all relevant mesh vertices, $k$, for multiple SOLPS-ITER cases. For all priors, the Cauchy distribution best encapsulated the prior argument distribution and the scale parameter, $\gamma$, was found.

Details of the spatially-independent priors are given in Table~\ref{tab:standard priors} and details of the spatially-dependent priors are given in Table~\ref{tab:grid-enabled priors}.

\begin{table*}[b]
    \begin{flushleft}
    \caption{
    The spatially-independent priors implemented within the IDA. When the argument deploys an upper and lower limit, the scale parameters are respectively listed.
    }
    \label{tab:standard priors}
    \end{flushleft}
    \centering
    \begin{tabular}{p{3cm}p{4.5cm}p{2cm}p{6cm}}
         \toprule
         Prior ($j$) & Argument ($f_j(\theta_k)$) & $\gamma_j$ & Justification \\
         \midrule 
         Static Electron Pressure,\newline $\quad P^k_e=T_e^k n_e^k$ & $\max\left(\frac{P^k_e}{P_e^{\mathrm{max}}} - 1, 0\right)$ & \standardform{1.5}{-2} & Static electron pressure is not expected to exceed $P_e^{max}=7\times10^{20}eVm^{-3}$ in MAST-U Super-X divertor, limit found through analysis of multiple SOLPS-ITER cases (Figure~4 of \cite{Bowman_2020}).\\
         Neutral Fraction,\newline $\quad f_{n_H}^k = \frac{n_H^k}{n_e^k + n_H^k}$ & $\max\left(\frac{f_0}{f_{n_H}^k}-1, \frac{f_{n_H}^k}{f(T_e)} -1, 0\right)$ & (\standardform{2.3}{-3},\newline\, \, \standardform{4.1}{-5}) & In MAST-U Super-X divertor, neutral fraction is not expected to exceed, $$f(T_e) = (1-c)\exp\left(-T_e^{k}/l\right)+c,$$ where
         $c=0.04$ and $l=5eV$. Similarly, the fraction was expected to remain above $f_0=0.002$.
         Limits found through analysis of multiple SOLPS-ITER cases (Figure~4 of \cite{Bowman_2020}). \\
         Helium ionisation ratio, \newline $\quad I_{He^{1+}}^k = \frac{n_{He^{1+}}^k}{n_{He^0}^k}$ & $\max\left(\frac{I_0(T_e)}{I_{He^{1+}}^k}-1, \frac{I_{He^{1+}}^k}{I_1(T_e)} -1, 0\right)$ & (\standardform{1.4}{-3},\newline\, \standardform{1.4}{-3}) & Limit set on the ratio of singly charged Helium density, $n_{He^{1+}}$ to neutral Helium density $n_{He_2}$. Temperature dependent limits, $I_0(T_e)$ and $I_1(T_e)$, empirically found through SOLPS-ITER case analysis.\\
         Helium concentration ratio, \newline $\quad C_{He}^k = \frac{n_{He^0}^k+n_{He^{1+}}^k}{n_e^k}$  & $\max\left(\frac{C_0}{C_{He}^k}-1, \frac{C_{He}^k}{C_1} -1, 0\right)$ & (\standardform{3.6}{-2},\newline\, \standardform{3.6}{-2}) & Limits on the expected ratio of helium density (neutral, $n_{He^0}$, and singly charger, $n_{He^{1+}}$, only) to the electron density in MAST-U Super-X divertor. Limits found empirically through analysis of SOLPS-ITER cases.\\ 
         Parameter bounds, \newline$\theta_F \in(T_e, n_e, n_H,Q_{mol.},$\newline $n_{He^{0}} n_{He^{1+}})$ & 
         $\max\left(\frac{\log(A)}{\log(\theta_F^k)} - 1, \frac{\log(\theta_F^k)}{\log(B)} - 1, 0\right)$ & - & $A\in($0.2, \standardform{1}{16}, \standardform{1}{15}, \standardform{1}{-3}, \standardform{1}{15}, \standardform{1}{13}$)$ respectively.\newline$B\in($60, \standardform{2.5}{20}, \standardform{2}{20}, \standardform{1}{3}, \standardform{2}{19}, \standardform{2}{19}$)$.\newline A normal distribution with $\sigma=$\standardform{1}{-4} was used in place of the Cauchy distribution. \\
         \bottomrule
    \end{tabular}
\end{table*}

\begin{table*}[b]
    \begin{flushleft}
    \caption{
    The spatially-dependent priors implemented within the IDA. Where multiple fields, F, are stated, the scale parameters are, respectively, listed. Derivatives $\partial_{s_{\parallel}}$, $\partial_{s_{\theta, \parallel}}$ and $\partial_{s_{\psi_N}}$ are with respect to distance parallel to total flux surfaces, distance parallel to surfaces of constant poloidal magnetic flux and distance perpendicular to surfaces of constant poloidal magnetic flux respectively. Gradients are approximated using finite difference with neighbouring mesh vertices.
    }
    \label{tab:grid-enabled priors}
    \end{flushleft}
    \centering
    \begin{tabular}{p{3cm}p{4cm}p{2cm}p{6.5cm}}
         \toprule
         Prior ($j$) & Argument ($f_j(\theta_k, ...)$) & $\gamma_j$ & Justification \\
         \midrule 
         Pressure Drop,\newline $\quad P^k_e=T_e^k n_e^k$ & 
         $\min\left(0, \frac{P^{k\mathrm{, upstream}}_e}{P_e^{k}} - 2\right)$\newline
         $\quad \cdot\;  \delta_{k\mathrm{\; at\; target}}$\newline
         $\quad \cdot\; \delta_{T_e^{k\mathrm{, upstream}}>5eV}$
         
         & (\standardform{1.7}{-2},\newline\, \standardform{1.7}{-2}) & From the two point model, expected minimally factor two reduction in static electron pressure from upstream to target along a flux tube\cite{stangeby2000plasma}. Since this pressure reduction argument loses validity if the upstream mesh vertex (`$k\mathrm{, upstream}$' is below the detachment threshold, this prior is only applied to flux tubes with an upstream electron temperature above 5eV. Prior is applied only at mesh vertices along the target.\\
         Parallel Gradient of log,\newline $\quad$ $f^{k}\in (T_e^{k}, T_e^k n_e^k)$ & $\max\left(0, \partial_{s_\parallel}F^k\right)$ & (\standardform{2.4}{-3},\newline\, \standardform{1.0}{-2}) & Expected monotonic reduction in electron temperature and static electron pressure on a path to the target, $s_\parallel$, parallel to surfaces of constant magnetic flux. \\
         Parallel Gradient of,\newline $\quad$ $Q^{k}_\parallel = -\kappa T_e^k\partial_{s_\parallel}T_e^k$ & $\max\left(0, \partial_{s_\parallel}Q^{k}_\parallel\right)$ & \standardform{4.0}{+2} & Expected monotonic reduction in parallel heat flux  on a path to the target, $s_\parallel$, parallel to surfaces of constant magnetic flux. The prior's argument form means the thermal conductivity, $\kappa$, can take any positive value.\\
         Perpendicular gradient of,\newline $\quad$ $F^k\in$\newline$(T_e^k, n_e^k, T_e^k n_e^k, n_{He^{1+}}^k)$ & $\max(0,$\newline  
         $\quad\quad\;\sgn\left( F^k - F^{k, \mathrm{max}} \right)$\newline $\quad\quad\;\cdot \partial_{s_{\psi_N}}F^k)$  & (\standardform{1.6}{-2},\newline\, \standardform{6.9}{-2},\newline\, \standardform{4.2}{-2},\newline\, \standardform{1.6}{-1}) & Expected single peak in electron temperature, electron density, static electron pressure and singly charged helium density on a path, $s_{\psi_N}$, perpendicular to surfaces of constant poloidal magnetic flux.\\
         Parallel smoothness of log,\newline $\quad$ $F^k\in(T_e^k, n_e^k, n_{He^{1+}}^k)$ & $\partial^2_{s_{\parallel} s_{\parallel}} \log\left(F^k\right)$ & (\standardform{1.2}{-1},\newline\, \standardform{1.0}{-1},\newline\, \standardform{1.7}{-1}) & Expected smoothness in the relative change of electron temperature, electron density and singly charged helium density along a path to the target, $s_\parallel$, parallel to surfaces of constant magnetic flux. \\
         Parallel smoothness of log,\newline $\quad$ $F^k\in(n_H^k, n_{He^0}^k)$ & $\partial^2_{s_{\theta, \parallel} s_{\theta, \parallel}} \log\left(F^k\right)$ & (\standardform{1.9}{+1},\newline\, \standardform{5.3}{+1}) & Expected smoothness in the relative change of neutral hydrogen and helium densities. Noting the lack of influence of magnetic fields on these species, we evaluate on our mesh on a path to the target, $s_{\theta, \parallel}$, parallel to surfaces of constant poloidal magnetic flux. \\
         Perpendicular smoothness\newline of log,\newline $\quad$ $F^k\in$\newline$(T_e^k, n_e^k, n_H^k, n_{He^0}^k, n_{He^{1+}}^k)$ & $\partial^2_{s_{\psi_N} s_{\psi_N}} \log\left(F^k\right)$ & (\standardform{1.8}{+2},\newline\, \standardform{2.2}{+2},\newline\, \standardform{3.3}{+2},\newline\, \standardform{7.9}{+2},\newline\, \standardform{7.6}{+2}) & Expected smoothness in the relative change of electron temperature and electron, neutral hydrogen, neutral helium and singly charged helium densities. We evaluate on our mesh on a path, $s_{\psi_N}$, perpendicular to surfaces of constant poloidal magnetic flux such that larger scale parameters can be applied in this direction.\\
         \bottomrule
    \end{tabular}
\end{table*}
\end{appendices}

\end{document}